\PassOptionsToPackage{table,xcdraw}{xcolor}
\documentclass[table,xcdraw,unnumsec,webpdf,contemporary,large]{oup-authoring-template}

\graphicspath{{Fig/}}

\usepackage{amsmath,amssymb,amsfonts}
\usepackage{multicol}
\usepackage{caption}
\usepackage{setspace}
\usepackage{textcomp}
\usepackage{mathtools}

\usepackage{xurl}
\usepackage{rotating}

\usepackage{hyperref}
\usepackage{natbib,stfloats}
\usepackage{mathrsfs}
\usepackage{lscape}

\usepackage{mdframed}
 
\theoremstyle{thmstyletwo}%
\theoremstyle{thmstylethree}%

\begin{document}

\journaltitle{Journal Title Here}
\DOI{DOI HERE}
\copyrightyear{2023}
\pubyear{2019}
\access{Advance Access Publication Date: Day Month Year}
\appnotes{Paper}

\firstpage{1}

\title[Identifying genes associated with phenotype]{Identifying genes associated with phenotypes using machine and deep learning}

\author[1,2]{Muhammad Muneeb}
\author[1,2,$\ast$]{David B. Ascher}
\author[1,2]{YooChan Myung}

\authormark{Muneeb et al.}

\address[1]{\orgdiv{School of Chemistry and Molecular Biology}, \orgname{The University of Queensland}, \orgaddress{\street{Queen Street}, \postcode{4067}, \state{Queensland}, \country{Australia}}}
\address[2]{\orgdiv{Computational Biology and Clinical Informatics}, \orgname{Baker Heart and Diabetes Institute}, \orgaddress{\street{Commercial Road}, \postcode{3004}, \state{Victoria}, \country{Australia}}}
 
\corresp[$\ast$]{Corresponding authors: David B. Ascher, Email: \href{email:d.ascher@uq.edu.au}{d.ascher@uq.edu.au}} 


\received{Date}{0}{Year}   
\revised{Date}{0}{Year}    
\accepted{Date}{0}{Year}   



\abstract{ 
Identifying disease-associated genes enables the development of precision medicine and the understanding of biological processes. Genome-wide association studies (GWAS), gene expression data, biological pathway analysis, and protein network analysis are among the techniques used to identify causal genes. We propose a machine-learning (ML) and deep-learning (DL) pipeline to identify genes associated with a phenotype. The proposed pipeline consists of two interrelated processes. The first is classifying people into case/control based on the genotype data. The second is calculating feature importance to identify genes associated with a particular phenotype. We considered 30 phenotypes from the openSNP data for analysis, 21 ML algorithms, and 80 DL algorithms and variants. The best-performing ML and DL models, evaluated by the area under the curve (AUC), F1 score, and Matthews correlation coefficient (MCC), were used to identify important single-nucleotide polymorphisms (SNPs), and the identified SNPs were compared with the phenotype-associated SNPs from the GWAS Catalog. The mean per-phenotype gene identification ratio (GIR) was 0.84. These results suggest that SNPs selected by ML/DL algorithms that maximize classification performance can help prioritise phenotype-associated SNPs and genes, potentially supporting downstream studies aimed at understanding disease mechanisms and identifying candidate therapeutic targets.}

\maketitle
\section{Introduction}
The interplay between genetic and environmental differences results in the variation in visual \cite{Jelenkovic2020} (height) and non-visual characteristics \cite{McGue1998} (behavior). Various environmental factors, such as diet, oxygen levels, and others, have been shown to influence many phenotypes directly \cite{Jaffee_2007,Williams_2000}. The degree to which these effects influence phenotypes is often influenced by subtle underlying genetic differences that can alter gene expression and biological processes \cite{Hunter2005}.  

The process of locating genes within a genome and determining their structure, function, and associations with specific traits or biological processes is known as gene identification \cite{Antonarakis2021,Fickett1996}. Several methods exist to identify genes associated with phenotypes, including genome-wide association studies (GWAS) \cite{Alqudah2020}, gene expression data analysis \cite{Roberts2022}, biological pathways analysis \cite{Cirillo2017}, protein network data analysis \cite{Wu2010}, transcriptome-wide association studies \cite{Li2021} and machine learning methods \cite{Mahendran2020,Pudjihartono2022}.
 
Traditionally, researchers use genome-wide association studies (GWAS) to identify SNPs that show strong associations with phenotypes. These association signals are then further investigated using approaches such as fine-mapping, gene expression analysis, colocalization studies, and biological pathway analysis to refine and interpret the genetic associations. In this study, we adopt a different strategy by using ML/DL models for variant prioritisation. Our approach assumes that models with higher classification performance identify SNPs that better discriminate cases from controls and are therefore more likely to be associated with the phenotype \cite{Musolf2021,Pudjihartono2022,Leal2019}. We first performed association testing and p-value thresholding to construct reduced SNP sets and then trained ML/DL models for phenotype classification. Feature-importance methods are subsequently used to rank SNPs and prioritise genes contributing most to phenotype prediction.
 
The process by which information encoded in a gene is used to produce a functional product, such as RNA or protein, is known as gene expression. Gene expression data are measured using technologies such as microarrays or RNA sequencing, and genes associated with specific phenotypes can be identified using statistical methods such as differential gene expression analysis \cite{SanSegundoVal2016,Ghadle2020,Maran2020,Anjum2016}. Gene expression data are used for gene and biomarker identification in diseases such as colorectal cancer \cite{Xu2020,Begum2021}, prostate cancer \cite{Lu2018}, brain cancer \cite{Senadheera2020}, bipolar disorder \cite{Liu2019}, and gastric cancer \cite{Kori2022} using bioinformatics frameworks. Additional experiments or data sources are typically required to confirm and interpret the biological relevance of the discovered genes, as gene expression data alone do not provide direct evidence of gene function \cite{Uygun2016}.

GWAS has been widely used to identify SNPs associated with phenotypes and involves scanning individual SNPs and comparing allele frequencies between cases and controls \cite{Uffelmann2021,CanoGamez2020}. The process involves recruiting people exhibiting variation in the phenotype of interest and sequencing the genotype data. The next step is to improve genotype data quality by removing SNPs with low genotype rates. After that, association testing is performed on each SNP to identify whether a particular SNP shows a significant difference in the allele frequency between cases and controls \cite{Marees2018}, and this methodology has been used for gene identification in diseases such as blood pressure \cite{Udosen2023}, Parkinson disease \cite{Kia2021}, and Alzheimer's disease \cite{Andrews2023}. Because the identified variants (SNPs) have limited predictive value, they cannot provide a comprehensive understanding of the underlying biological mechanisms.

Machine learning methods like Random Forest and Gradient Boosting \cite{Gill2022}, Gradient Boosting Machine \cite{Li2018}, and Support Vector Machine (SVM) \cite{Gaudillo2019} can be applied to association studies to evaluate the significance of SNPs in predicting phenotype status \cite{Mahendran2020,Tabl2019}. These methods assess the usefulness of each SNP, and the SNP significance score can be used to order the SNPs and identify those of high importance \cite{Musolf2021}. These methods use permutation-based \cite{Mi2021} and feature importance techniques to rank SNPs and have been applied to breast cancer \cite{Tabl2019}, pancreatic cancer \cite{Mori2021}, cancer \cite{Gomes2022}, and Alzheimer's disease \cite{Alatrany2023}. Moreover, feature selection approaches such as recursive feature elimination and the Least Absolute Shrinkage and Selection Operator (LASSO) regression can also be used to rank SNPs.

Machine learning algorithms have been used for gene function prediction \cite{Mahood2020}, gene-gene interactions \cite{McKinney2006}, and gene expression analysis \cite{Abbas2020}. In this study, we used machine learning to identify genes from genotype data.

Among methods that use genotype data for gene identification, GWAS and machine/deep learning are prominent, relying on SNPs to identify associated genes. Every gene is composed of multiple SNPs; identifying SNPs associated with a particular phenotype or trait results in identifying a gene associated with a phenotype \cite{Alqudah2020}. Some SNP variations can modify gene expression, protein structure, or protein-protein interactions, thereby affecting functional outcomes and contributing to complex phenotypic features, including disease susceptibility and drug responsiveness, making them good candidates for gene identification. 

We used ML and DL algorithms for genotype-phenotype prediction. Then, the best-performing model was used to rank the SNPs or features that contributed most to the separation of cases and controls, and the identified SNPs for each phenotype were compared with existing phenotype-associated SNPs from the GWAS Catalog. The proposed methodology was tested on 30 phenotypes from the openSNP data, and ML/DL algorithms identified genes (listed in the GWAS Catalog) associated with phenotypes.

\section{Methodology}
\label{methodology}
\subsection{Modeling}
\begin{figure*}[!ht]
\centering
 \includegraphics[width=11cm,height=14cm]{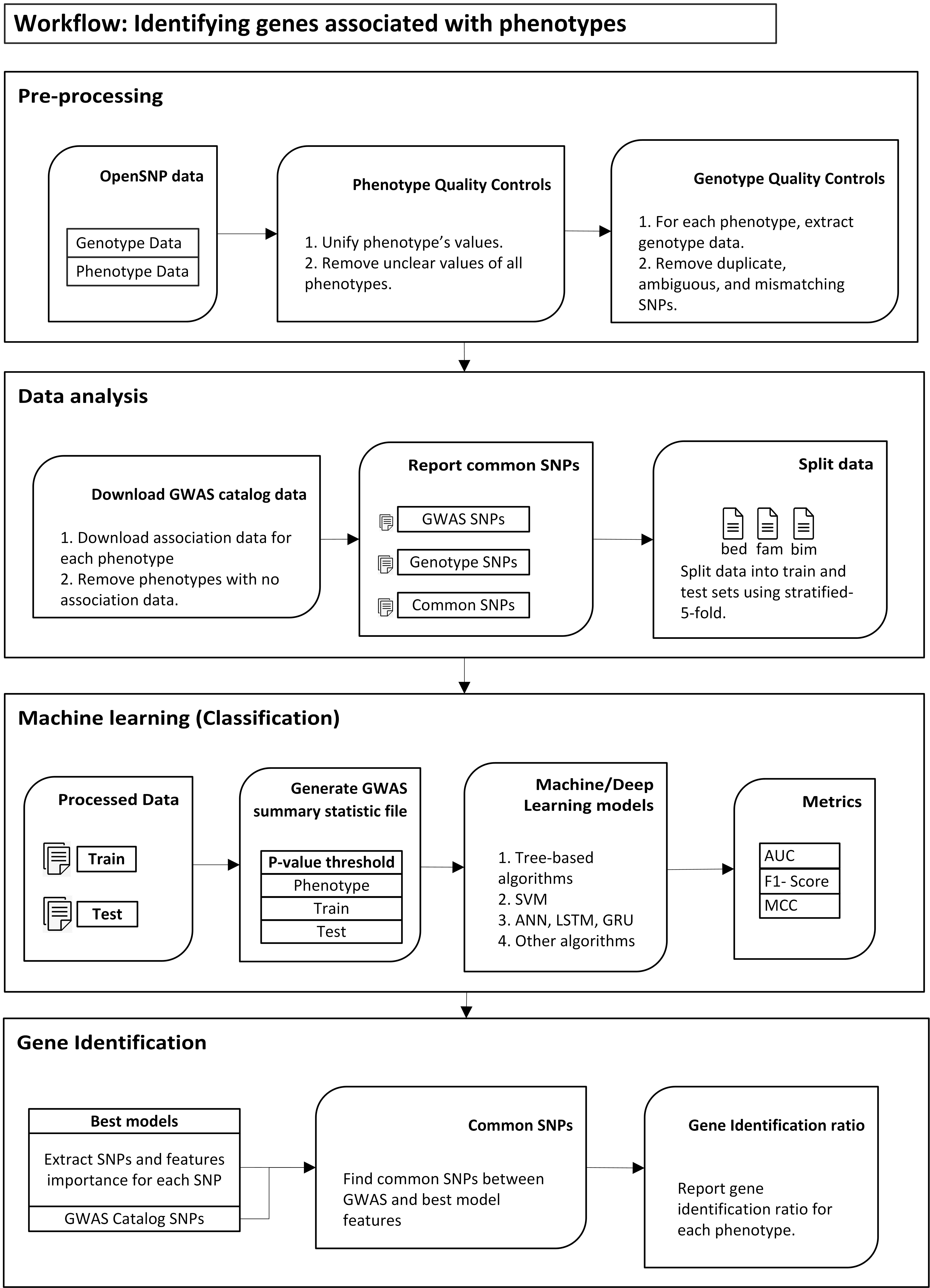}
\caption{\textbf{A workflow of identifying genes associated with phenotypes.} First, clean phenotype data, convert genotype data to PLINK format, and perform quality control steps on genotype data. Then, download gene associations for each phenotype. Split the data into five folds and generate bed, bim, and fam files for each split. Divide genotype data into training and test sets, generate sub-datasets using p-value thresholds, pass data to machine/deep learning models, and report AUC, MCC, and F1-Score for each model and p-value thresholds. Finally, list the feature (SNPs) importance for the best-performing models, compare the identified SNPs with those downloaded from the GWAS Catalog, and report the gene identification ratio for each phenotype.}
\label{flowchart}
\end{figure*}

We considered binary phenotypes from the openSNP data \cite{Greshake2014}, and the dataset processing steps are explained in detail in the supplementary material (section 1). The following describes the process after genotype data conversion. Figure \ref{flowchart} shows the overall workflow of identifying genes associated with phenotypes using machine learning models. 

Duplicate genotype files and SNPs were discarded, leaving 80 phenotypes. A list is available on GitHub (Analysis3.pdf). We considered a Hardy--Weinberg equilibrium threshold of $1\times10^{-6}$, a genotype missingness threshold of $0.01$, a minor allele frequency threshold of $0.01$, and an individual missingness threshold of $0.7$ \cite{McCarthy2008,Clayton2005} to improve genotype data quality.


We searched the GWAS Catalog for each phenotype and found SNP association data available for 36 phenotypes. The table showing the association study number for each phenotype is available on GitHub (Table1.md). Each phenotype has multiple association files, and we considered one association file with a high name resemblance to a specific phenotype. We downloaded the associations for each phenotype and checked whether these SNPs were present in the processed genotype data. Table \ref{SNPscount} lists the phenotype, the number of associated SNPs from the GWAS Catalog, the number of SNPs in the genotype data for a particular phenotype, and the number of common SNPs between the GWAS Catalog and genotype data. Six phenotypes with no common SNPs were removed, leaving 30 for further analysis.

\begin{table*}[!ht]
\centering
\resizebox{1.5\columnwidth}{!}{%
\begin{tabular}{|l|l|l|l|}
\hline
\textbf{Phenotype}                & \textbf{SNPs in GWAS Catalog} & \textbf{SNPs in our data} & \textbf{Common SNPs} \\ \hline
Attention Deficit Hyperactivity Disorder (ADHD) & 1818 & 63900  & 42  \\ \hline
Allergic rhinitis               & 399  & 102988 & 16  \\ \hline
Amblyopia                      & 15   & 101302 & 0   \\ \hline
Asthma                         & 3291 & 681937 & 408 \\ \hline
Bipolar Disorder                & 1585 & 101478 & 66  \\ \hline
Cholesterol                    & 5000 & 69425  & 130 \\ \hline
Cluster headache                & 6    & 102495 & 0   \\ \hline
Craves sugar                    & 85   & 104592 & 2   \\ \hline
Dental decay                    & 294  & 105383 & 24  \\ \hline
Depression                     & 2586 & 99857  & 82  \\ \hline
Diagnosed Vitamin D deficiency    & 30   & 105035 & 4   \\ \hline
Diagnosed with Sleep Apnea        & 73   & 101361 & 6   \\ \hline
Dyslexia                       & 126  & 104630 & 2   \\ \hline
Earlobe Free or attached          & 201  & 100852 & 11  \\ \hline
Eczema                         & 505  & 104091 & 15  \\ \hline
Generalized Anxiety Disorder     & 23   & 67828  & 0   \\ \hline
Hair Type                       & 21   & 687752 & 7   \\ \hline
Have Myalgic Encephalomyelitis/Chronic Fatigue Syndrome (MECFS) & 5 & 106221 & 0 \\ \hline
Hypertension                   & 688  & 73859  & 52  \\ \hline
Hypertriglyceridemia           & 36   & 218086 & 4   \\ \hline
Irritable Bowel Syndrome         & 34   & 105698 & 2   \\ \hline
Mental Disease                  & 4988 & 98153  & 180 \\ \hline
Migraine                       & 642  & 107740 & 25  \\ \hline
Motion Sickness                 & 35   & 71798  & 3   \\ \hline
Panic Disorder                  & 24   & 105010 & 2   \\ \hline
Photic sneeze reflex (photoptarmic reflex) & 56   & 97380  & 4   \\ \hline
Plantar fasciitis               & 2    & 106492 & 0   \\ \hline
Post-Traumatic Stress Disorder (PTSD) & 137 & 102913 & 5 \\ \hline
Restless leg syndrome            & 64   & 100985 & 6   \\ \hline
Scoliosis                      & 1392 & 100822 & 29  \\ \hline
Seborrhoeic Dermatitis          & 12   & 105299 & 0   \\ \hline
Sensitivity to Mosquito Bites     & 434  & 103606 & 3   \\ \hline
Sleep Disorders                 & 4067 & 101291 & 93  \\ \hline
Strabismus                     & 30   & 101765 & 1   \\ \hline
Thyroid Issues Cancer            & 92   & 101747 & 8   \\ \hline
TypeIIDiabetes                 & 5000 & 100694 & 193 \\ \hline
\end{tabular}%
}
\caption{\textbf{Phenotypes and the number of overlapping SNPs between the genotype data and the GWAS Catalog.} 
The table lists phenotypes and the number of SNPs shared between the GWAS Catalog and the genotype dataset. No overlapping SNPs were identified for six phenotypes (Amblyopia, Cluster headache, Generalized anxiety disorder, ME/CFS, Plantar fasciitis, and Seborrhoeic dermatitis); therefore, these phenotypes were excluded from further analysis.}
\label{SNPscount}
\end{table*}


We used PLINK to split the data (stratified \texttt{5}-fold) into training (80\%) and test sets (20\%), and we performed the following steps on each fold. The training data were subjected to Fisher's exact test for allelic association (using \texttt{plink --assoc}) to generate the GWAS summary statistics, which contain p-values for each SNP. Using p-value thresholding, we extracted the top 50, 100, 200, 500, 1000, 5000, and 10000 SNPs from the training and test data and saved the encoded data in a raw tabular format for model training.


We considered multiple hyperparameters for the deep learning models, while we used the default hyperparameters for the machine learning models. We adjusted the hyperparameters of the deep learning models to obtain a new version of the model. We used AUC, F1 score, and MCC to assess the model's performance. We averaged the AUC, F1 score, and MCC across all folds for each model to identify the best-performing model for each evaluation metric \cite{Liao2022}.






We listed the features from the models yielding the best AUC, F1 Score, and MCC for each phenotype. A single model that yields the best performance in terms of AUC may not necessarily yield the best performance in terms of MCC or F1 Score \cite{Hicks2022}, and selecting multiple models optimized for different evaluation metrics helps identify the discriminative boundary between cases and controls and improves feature selection. We extracted the top-ranked SNPs from each model's feature-importance method and compared them with SNPs and genes from the GWAS Catalog.

\subsection{Implementation}
This section provides implementation details of the ML/DL models and the feature importance process for both techniques, which affect gene identification.


\subsubsection{Machine learning - models}
We used 21 machine learning algorithms and their variants implemented in the \texttt{scikit-learn} library \cite{scikit-learn}. The algorithms include tree-based classification algorithms \cite{asd} (AdaBoost, XGBoost \cite{Chen:2016:XST:2939672.2939785}, Random Forest, and Gradient boosting algorithms), Stochastic Gradient Descent (SGD), Support Vector Classifier (SVC) \cite{Cristianini2008}, and other algorithms. The list of 21 machine-learning algorithms is available on GitHub (MachineLearningAlgorithms.txt).

To calculate feature importance for SVC and similar methods, extract the coefficients of the learned hyperplane, take the absolute values of the coefficients, normalize the coefficient values, and those values are feature importance for all the features used to train the machine learning algorithm \cite{Saarela2021}. Tree-based algorithms, such as Random Forest, compute feature importance by measuring how much impurity is reduced by splitting each feature. XGBoost computes feature importance as the number of times a feature is used in the boosting process and the magnitude of its contribution to the loss function \cite{Pudjihartono2022,Chen2020}.



\subsubsection{Deep learning - models} 
We used four deep learning models: Artificial neural network (ANN) \cite{mcculloch1943logical}, Gated recurrent unit (GRU) \cite{8053243}, Long Short-Term Memory (LSTM) \cite{Hochreiter1997}, and Bidirectional LSTM (BILSTM) \cite{Schuster1997}. There were five layers in each model, and the numbers of neurons in each layer were 128 + $\sqrt[2]{S}$, 64 + $\sqrt[4]{S}$, 32 + $\sqrt[8]{S}$, 16 + $\sqrt[16]{S}$, and 1, where S is the number of SNPs in the genotype data. As genotype data contain different numbers of SNPs, the models adjust themselves to the input size, and using the root of the number of SNPs allows the same model to be trained on datasets of different dimensionality. We created six new stacked model architectures using GRU, LSTM, and BILSTM layers. We generated 80 deep learning models by varying four hyperparameters, as shown in Table \ref{deeplearningparameters}. This combination of hyperparameters yielded 8 distinct models of a specific architecture. The list of 80 deep learning algorithms is available on GitHub (DeepLearningAlgorithms.txt).

\begin{table}[!ht]
\begin{tabular}{|l|l|}
\hline
\textbf{Parameters}    & \textbf{Values} \\ \hline
\textbf{Dropout}       & 0.2,0.5         \\ \hline
\textbf{Optimizer}     & Adam            \\ \hline
\textbf{Batch size}    & 1,5             \\ \hline
\textbf{Epochs number} & 50,200          \\ \hline
\end{tabular}
      \caption{\textbf{Deep learning model's hyperparameters and values.} The first column shows the deep learning model's hyperparameters, and the second column shows the values we considered for each hyperparameter. The combination of hyperparameter values yielded 8 $(2*1*2*2)$ models for a given architecture.}
\label{deeplearningparameters}
\end{table}

We used feature dropout to compute feature importance. First, train the model, evaluate baseline performance, iterate over each input feature individually, drop that feature, evaluate performance, record the change in performance relative to the baseline, and rank the dropped feature by the change in performance. A larger drop in performance suggests higher importance, while a small drop implies lower importance \cite{FigueroaBarraza2021,cai2018feature,NEURIPS2020}.

\begin{figure}[!ht]
\centering
 \includegraphics[width=8cm,height=3cm]{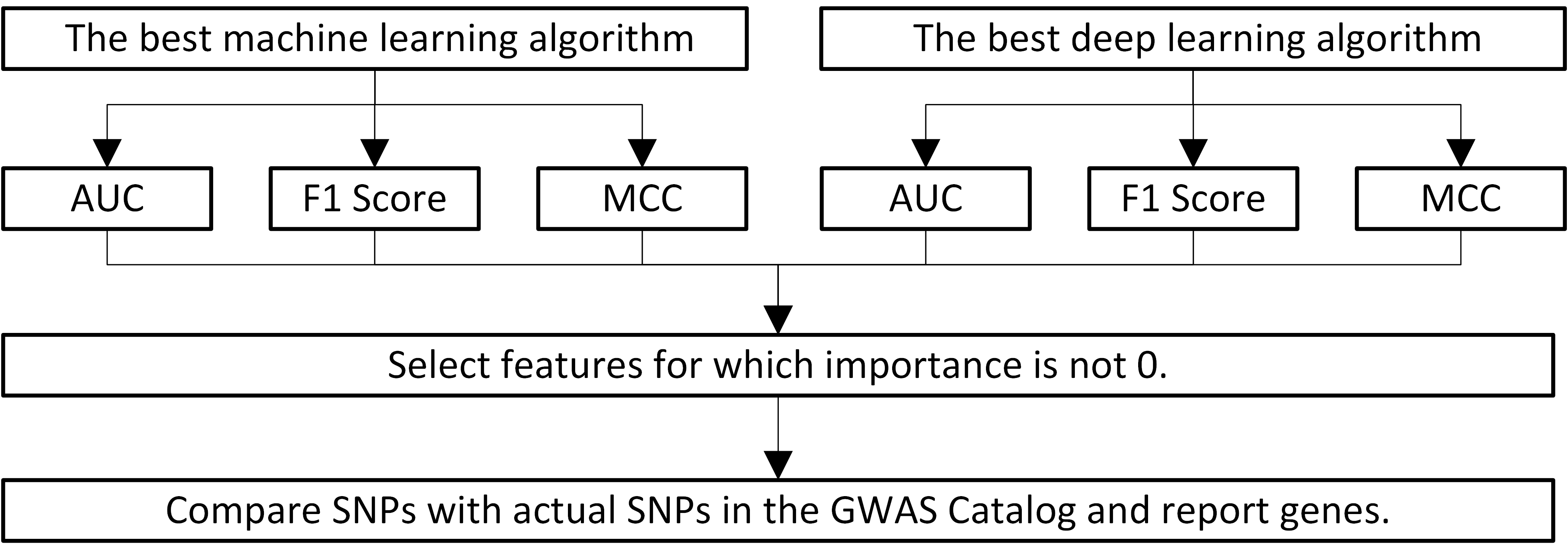}
\caption{\textbf{A process of listing features identified by machine and deep learning models.} We extracted the top-ranked SNPs from the best-performing machine and deep learning models in terms of AUC, F1 Score, and MCC. Those SNPs were compared with the actual SNPs from the GWAS Catalog.}
\label{heatmap22}
\end{figure}

Once we identified all models that yielded the best performance on AUC, MCC, and F1 Score for a specific phenotype, we ranked SNPs by feature-dropout importance and compared the highest-ranked SNPs with GWAS Catalog associations. The process is explained in Figure \ref{heatmap22}.

\section{Results}

\subsection{Classification performance of machine and deep learning models}
\begin{figure*}[!ht]
\centering
\includegraphics[width=15cm,height=17cm]{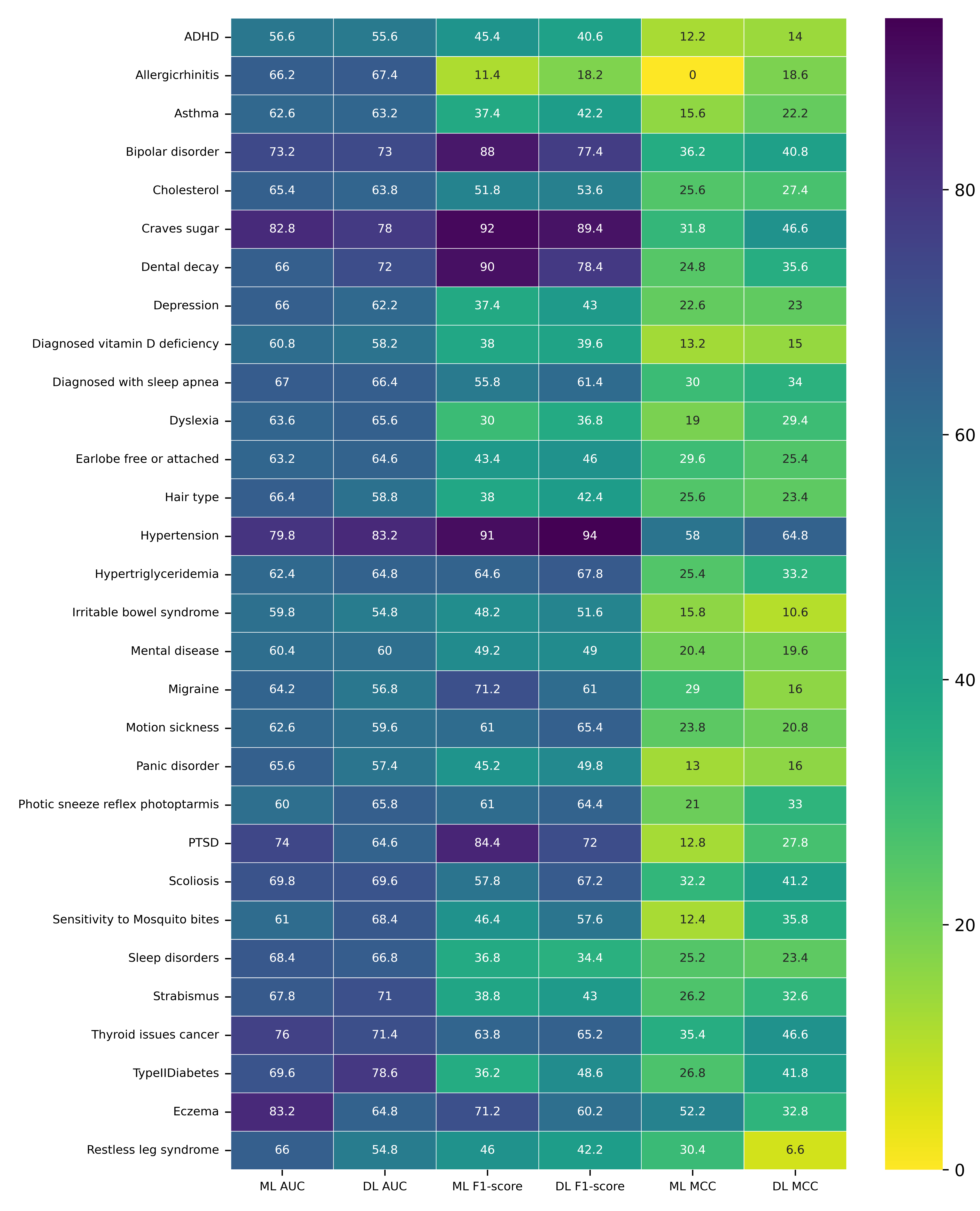}
\caption{\textbf{A heatmap showing the AUC, F1 score, and MCC for machine and deep learning.} This heatmap shows the classification results from the best-performing models from the deep \texttt{(DL)} and machine \texttt{(ML)} learning algorithms in terms of AUC, F1 score, and MCC. }
\label{heatmap}
\end{figure*}

Figure \ref{heatmap} shows the classification results for all phenotypes. The detailed results for the best-performing models, in terms of AUC, F1 Score, and MCC, for both ML and DL are provided in Supplementary Section 1.

Among machine learning algorithms, XGBoost and its variants achieved the highest AUCs, yielding the best results for 18 phenotypes. The SGD classifier was the best-performing model for 15 phenotypes in terms of MCC and for 9 phenotypes in terms of F1 Score. Among deep learning algorithms, ANN performed best for most phenotypes across all evaluation metrics. 

We calculated the average performance of ML/DL algorithms in terms of AUC, F1 Score, and MCC across all phenotypes. We found that deep learning algorithms achieved better performance on MCC and F1 Score, whereas machine learning achieved better performance on AUC, as shown in Table \ref{machinevsdeepclassification}.

\begin{table}[!ht]
\begin{tabular}{|l|l|l|}
\hline
 & \textbf{Machine learning} & \textbf{Deep learning} \\ \hline
\textbf{AUC} & 0.65 & 0.63 \\ \hline
\textbf{F1 Score} & 0.53 & 0.54 \\ \hline
\textbf{MCC} & 0.24 & 0.28 \\ \hline
\end{tabular}
\caption{\textbf{Average classification performance of machine- and deep-learning algorithms in terms of AUC, F1 score, and MCC.} Performance for each metric was calculated as the mean across phenotypes, where each phenotype metric was averaged across cross-validation folds.}
\label{machinevsdeepclassification}
\end{table}

\subsection{Gene Identification}
\begin{figure*}[!ht]
\centering
 \includegraphics[width=18cm,height=15cm]{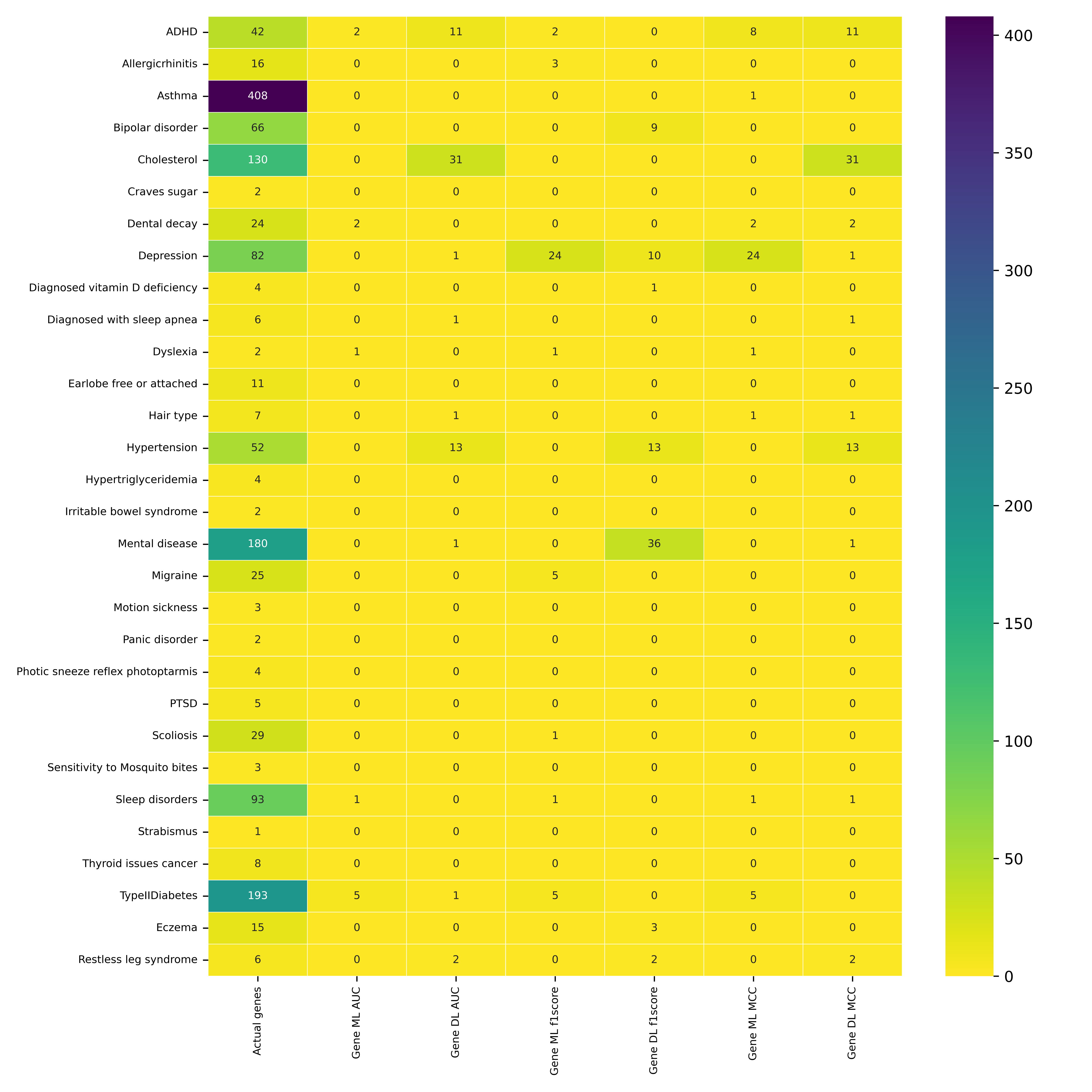}
\caption{\textbf{Heatmap of genes identified by the best-performing ML/DL models according to AUC, F1 score, and MCC.} The first column shows the number of SNPs or genes overlapping between the genotype dataset and the GWAS Catalog. ML and DL denote machine learning and deep learning, respectively. Columns 2--7 show the number of SNPs identified by ML and DL models that achieved the best performance for AUC, F1 score, and MCC.}
\label{genesperformance}
\end{figure*}

We investigated the number of causal genes identified by the best-performing models. Figure \ref{genesperformance} shows the number of genes identified for each phenotype when the best-performing ML/DL algorithms (in terms of AUC, F1 Score, and MCC) were used to rank SNPs based on feature importance. 

We calculated correlations between the best classification performance achieved by ML and DL models, measured using AUC, F1 score, and MCC, and the number of genes identified by the algorithms for each phenotype (Table \ref{performanceandgenecorrelation}). This analysis evaluates whether a particular model type (ML or DL) and evaluation metric (AUC, F1 score, or MCC) is associated with identifying a greater number of genes. The results indicate that deep-learning models optimized for MCC show a relatively stronger positive correlation, suggesting that optimizing MCC may be associated with identifying more genes than optimizing AUC or F1 score.

\begin{table}[!ht]
\begin{tabular}{|l|l|l|}
\hline
 & \textbf{Machine learning} & \textbf{Deep learning} \\ \hline
\textbf{AUC} & 0.02 & 0.06 \\ \hline
\textbf{F1 Score} & -0.18 & 0.1 \\ \hline
\textbf{MCC} & -0.08 & 0.11 \\ \hline
\end{tabular}
\caption{A table showing the correlation between the best-performing model's classification score and the number of genes identified.}
\label{performanceandgenecorrelation}
\end{table}

From the results, we observed three cases of gene identification: first, no gene was identified; second, there was a high correlation between high classification performance and the number of genes identified; third, genes were identified by either the machine or deep learning algorithm, independent of performance.

For 11 phenotypes, no common gene was identified by either the ML/DL algorithms, even though both achieved high classification performance (colored grey in Table \ref{finaltable}). There are four potential reasons why no gene was identified despite a good performance by a machine or deep learning algorithm: (1) genotype data quality, with low coverage SNPs discarded by the algorithms; (2) linkage disequilibrium can remove highly linked SNPs; (3) the non-linear nature of the machine and deep learning algorithm, leading to combinations of unrelated SNPs being given more weight than causative SNPs \cite{Roy2021}; or (4) population structure, as the reported causative genes on the GWAS Catalog may come from different population distributions than the sample population \cite{Barreiro2008}.

In the second gene identification scenario, there was a strong correlation between the number of genes identified and the algorithm's performance. For nine phenotypes (shown in green in Table \ref{finaltable}), the machine and deep learning algorithms exhibited robust, generalizable performance, resulting in more causative genes being identified. This aligns with the assumption that optimal algorithms would identify SNPs that yield a better separation boundary between cases and controls. The gene identification ratio (GIR; Equation \ref{geneidentificationratio}) for most of these phenotypes was between 0.1 and 0.3.

\begin{multline}
    \text{Gene Identification Ratio} = \\
    \frac{\text{Number of Genes Identified by Machine/Deep Learning}}{\text{Number of genes from the GWAS Catalog}}
    \label{geneidentificationratio}
\end{multline}

In the third gene identification case, the machine or deep learning algorithm identified the causative genes, but this was not linked to performance metrics (as shown in yellow in Table \ref{finaltable}). The reason is that ML/DL algorithms use different classification methods. To address this issue, we considered combining the genes identified by both methods and developed an ensemble approach to improve gene identification, as we have done in this study.

We further investigated the gene identification results to identify the reasons for the three cases reported in Table \ref{finaltable}.

We applied p-value thresholds to reduce the number of SNPs and used those SNPs for model training. The purpose is to determine the minimum number of SNPs that maximizes classification performance. When SNPs are reduced, the actual number of common SNPs between the data and the GWAS Catalog is reduced. We counted the common SNPs from the best-performing p-value threshold and the GWAS Catalog for each phenotype to further explore this (see Table \ref{newgeneidentificationratio}, column \texttt{(A2)}). 

After this process, we calculated the gene identification ratio for each phenotype (see Table \ref{newgeneidentificationratio}, column \texttt{(New GIR)}), and it changed the results, highlighting that common SNPs have been reduced. The discarded SNPs do not appear in the p-value threshold applied to the GWAS summary statistic file, suggesting that they are less associated with the phenotype under consideration in our dataset.

Table \ref{newgeneidentificationratio} shows the new gene identification ratio for each phenotype. Overall, the mean of the per-phenotype gene identification ratios was 0.84; however, phenotypes with very few GWAS-overlapping genes can disproportionately inflate or deflate this average, so it should be interpreted alongside the individual ratios in Table \ref{newgeneidentificationratio}.

We found that individual missingness impacts gene identification performance. We calculated missingness per individual for each phenotype using PLINK. We found a statistically non-significant inverse correlation of -0.15 between the number of genes identified and the missingness per individual, suggesting that higher genotype rates are associated with more common genes identified by machine/deep learning.

\begin{landscape}
\begin{table}[]
\begin{tabular}{|l|l|l|l|l|l|l|l|l|l|l|l|l|l|l|}
\hline
\textbf{Phenotypes} & \textbf{CMA} & \textbf{GMA} & \textbf{CMF} & \textbf{GMF} & \textbf{CMM} & \textbf{GMM} & \textbf{CDA} & \textbf{GDA} & \textbf{CDF} & \textbf{GDF} & \textbf{CDM} & \textbf{GDM} & \textbf{I} & \textbf{A} \\ \hline
\rowcolor[HTML]{FCFF2F} 
\textbf{ADHD} & 56.6 & 2 & 45.4 & 2 & 12.2 & 8 & 55.6 & 11 & 40.6 & 0 & 14 & 11 & 13 & 42 \\ \hline
\rowcolor[HTML]{FCFF2F} 
\textbf{Allergic rhinitis} & 66.2 & 0 & 11.4 & 3 & 0 & 0 & 67.4 & 0 & 18.2 & 0 & 18.6 & 0 & 3 & 16 \\ \hline
\rowcolor[HTML]{FCFF2F} 
\textbf{Asthma} & 62.6 & 0 & 37.4 & 0 & 15.6 & 1 & 63.2 & 0 & 42.2 & 0 & 22.2 & 0 & 1 & 408 \\ \hline
\rowcolor[HTML]{FCFF2F} 
\textbf{Bipolar disorder} & 73.2 & 0 & 88 & 0 & 36.2 & 0 & 73 & 0 & 77.4 & 9 & 40.8 & 0 & 9 & 66 \\ \hline
\rowcolor[HTML]{67FD9A} 
\textbf{Cholesterol} & 65.4 & 0 & 51.8 & 0 & 25.6 & 0 & 63.8 & 31 & 53.6 & 0 & 27.4 & 31 & 31 & 130 \\ \hline
\rowcolor[HTML]{C0C0C0} 
\textbf{Craves sugar} & 82.8 & 0 & 92 & 0 & 31.8 & 0 & 78 & 0 & 89.4 & 0 & 46.6 & 0 & 0 & 2 \\ \hline
\rowcolor[HTML]{67FD9A} 
\textbf{Dental decay} & 66 & 2 & 90 & 0 & 24.8 & 2 & 72 & 0 & 78.4 & 0 & 35.6 & 2 & 2 & 24 \\ \hline
\rowcolor[HTML]{FCFF2F} 
\textbf{Depression} & 66 & 0 & 37.4 & 24 & 22.6 & 24 & 62.2 & 1 & 43 & 10 & 23 & 1 & 24 & 82 \\ \hline
\rowcolor[HTML]{67FD9A} 
\textbf{Diagnosed vitamin D deficiency} & 60.8 & 0 & 38 & 0 & 13.2 & 0 & 58.2 & 0 & 39.6 & 1 & 15 & 0 & 1 & 4 \\ \hline
\rowcolor[HTML]{67FD9A} 
\textbf{Diagnosed with sleep apnea} & 67 & 0 & 55.8 & 0 & 30 & 0 & 66.4 & 1 & 61.4 & 0 & 34 & 1 & 1 & 6 \\ \hline
\rowcolor[HTML]{FCFF2F} 
\textbf{Dyslexia} & 63.6 & 1 & 30 & 1 & 19 & 1 & 65.6 & 0 & 36.8 & 0 & 29.4 & 0 & 1 & 2 \\ \hline
\rowcolor[HTML]{C0C0C0} 
\textbf{Earlobe free or attached} & 63.2 & 0 & 43.4 & 0 & 29.6 & 0 & 64.6 & 0 & 46 & 0 & 25.4 & 0 & 0 & 11 \\ \hline
\rowcolor[HTML]{FCFF2F} 
\textbf{Hair type} & 66.4 & 0 & 38 & 0 & 25.6 & 1 & 58.8 & 1 & 42.4 & 0 & 23.4 & 1 & 1 & 7 \\ \hline
\rowcolor[HTML]{67FD9A} 
\textbf{Hypertension} & 79.8 & 0 & 91 & 0 & 58 & 0 & 83.2 & 13 & 94 & 13 & 64.8 & 13 & 13 & 52 \\ \hline
\rowcolor[HTML]{C0C0C0} 
\textbf{Hypertriglyceridemia} & 62.4 & 0 & 64.6 & 0 & 25.4 & 0 & 64.8 & 0 & 67.8 & 0 & 33.2 & 0 & 0 & 4 \\ \hline
\rowcolor[HTML]{C0C0C0} 
\textbf{Irritable bowel syndrome} & 59.8 & 0 & 48.2 & 0 & 15.8 & 0 & 54.8 & 0 & 51.6 & 0 & 10.6 & 0 & 0 & 2 \\ \hline
\rowcolor[HTML]{67FD9A} 
\textbf{Mental disease} & 60.4 & 0 & 49.2 & 0 & 20.4 & 0 & 60 & 1 & 49 & 36 & 19.6 & 1 & 36 & 180 \\ \hline
\rowcolor[HTML]{67FD9A} 
\textbf{Migraine} & 64.2 & 0 & 71.2 & 5 & 29 & 0 & 56.8 & 0 & 61 & 0 & 16 & 0 & 5 & 25 \\ \hline
\rowcolor[HTML]{C0C0C0} 
\textbf{Motion sickness} & 62.6 & 0 & 61 & 0 & 23.8 & 0 & 59.6 & 0 & 65.4 & 0 & 20.8 & 0 & 0 & 3 \\ \hline
\rowcolor[HTML]{C0C0C0} 
\textbf{Panic disorder} & 65.6 & 0 & 45.2 & 0 & 13 & 0 & 57.4 & 0 & 49.8 & 0 & 16 & 0 & 0 & 2 \\ \hline
\rowcolor[HTML]{C0C0C0} 
\textbf{Photic sneeze reflex (photoptarmic reflex)} & 60 & 0 & 61 & 0 & 21 & 0 & 65.8 & 0 & 64.4 & 0 & 33 & 0 & 0 & 4 \\ \hline
\rowcolor[HTML]{C0C0C0} 
\textbf{PTSD} & 74 & 0 & 84.4 & 0 & 12.8 & 0 & 64.6 & 0 & 72 & 0 & 27.8 & 0 & 0 & 5 \\ \hline
\rowcolor[HTML]{67FD9A} 
\textbf{Scoliosis} & 69.8 & 0 & 57.8 & 1 & 32.2 & 0 & 69.6 & 0 & 67.2 & 0 & 41.2 & 0 & 1 & 29 \\ \hline
\rowcolor[HTML]{C0C0C0} 
\textbf{Sensitivity to Mosquito bites} & 61 & 0 & 46.4 & 0 & 12.4 & 0 & 68.4 & 0 & 57.6 & 0 & 35.8 & 0 & 0 & 3 \\ \hline
\rowcolor[HTML]{67FD9A} 
\textbf{Sleep disorders} & 68.4 & 1 & 36.8 & 1 & 25.2 & 1 & 66.8 & 0 & 34.4 & 0 & 23.4 & 1 & 1 & 93 \\ \hline
\rowcolor[HTML]{C0C0C0} 
\textbf{Strabismus} & 67.8 & 0 & 38.8 & 0 & 26.2 & 0 & 71 & 0 & 43 & 0 & 32.6 & 0 & 0 & 1 \\ \hline
\rowcolor[HTML]{C0C0C0} 
\textbf{Thyroid issues cancer} & 76 & 0 & 63.8 & 0 & 35.4 & 0 & 71.4 & 0 & 65.2 & 0 & 46.6 & 0 & 0 & 8 \\ \hline
\rowcolor[HTML]{FCFF2F} 
\textbf{TypeIIDiabetes} & 69.6 & 5 & 36.2 & 5 & 26.8 & 5 & 78.6 & 1 & 48.6 & 0 & 41.8 & 0 & 5 & 193 \\ \hline
\rowcolor[HTML]{FCFF2F} 
\textbf{Eczema} & 83.2 & 0 & 71.2 & 0 & 52.2 & 0 & 64.8 & 0 & 60.2 & 3 & 32.8 & 0 & 3 & 15 \\ \hline
\rowcolor[HTML]{FCFF2F} 
\textbf{Restless leg syndrome} & 66 & 0 & 46 & 0 & 30.4 & 0 & 54.8 & 2 & 42.2 & 2 & 6.6 & 2 & 2 & 6 \\ \hline
\end{tabular}
\caption{\textbf{A table showing the AUC, MCC, and F1 Score and the number of genes identified by each model.}
In this table, labels for the first 12 columns are shown in this format \texttt{(XYZ)}. \texttt{X} can be \texttt{[C, G]}, where \texttt{C} = the classification performance and \texttt{G} = the number of genes identified. \texttt{Y} can be \texttt{[M, D]}, where \texttt{M} means machine learning and \texttt{D} means deep learning. \texttt{Z} can be \texttt{[A, F, M]}, where \texttt{A = AUC}, \texttt{F = F1 score}, and \texttt{M = MCC}. \texttt{I} means the total number of genes identified by the machine and deep learning algorithm combined. The last column \texttt{A} shows the number of genes associated with the phenotype. There are three color codes, green, yellow, and gray, which are explained in the text.
}
\label{finaltable}
\end{table}
\end{landscape}

\begin{table*}[!ht]
\begin{tabular}{|
>{\columncolor[HTML]{FFFFFF}}l |
>{\columncolor[HTML]{FFFFFF}}l |
>{\columncolor[HTML]{FFFFFF}}l |
>{\columncolor[HTML]{FFFFFF}}l |
>{\columncolor[HTML]{FFFFFF}}l |l|}
\hline
\textbf{Phenotypes} & \textbf{A} & \textbf{I} & \textbf{GIR} & \textbf{A2} & \textbf{New GIR} \\ \hline
{\color[HTML]{333333} \textbf{ADHD}} & {\color[HTML]{333333} 42} & {\color[HTML]{333333} 13} & {\color[HTML]{333333} 0.31} & {\color[HTML]{333333} 14} & 0.93 \\ \hline
{\color[HTML]{333333} \textbf{Allergic rhinitis}} & {\color[HTML]{333333} 16} & {\color[HTML]{333333} 3} & {\color[HTML]{333333} 0.19} & {\color[HTML]{333333} 3} & 1 \\ \hline
{\color[HTML]{333333} \textbf{Asthma}} & {\color[HTML]{333333} 408} & {\color[HTML]{333333} 1} & {\color[HTML]{333333} 0.0} & {\color[HTML]{333333} 1} & 1 \\ \hline
{\color[HTML]{333333} \textbf{Bipolar disorder}} & {\color[HTML]{333333} 66} & {\color[HTML]{333333} 9} & {\color[HTML]{333333} 0.14} & {\color[HTML]{333333} 13} & 0.69 \\ \hline
{\color[HTML]{333333} \textbf{Cholesterol}} & {\color[HTML]{333333} 130} & {\color[HTML]{333333} 31} & {\color[HTML]{333333} 0.24} & {\color[HTML]{333333} 38} & 0.81 \\ \hline
{\color[HTML]{333333} \textbf{Craves sugar}} & {\color[HTML]{333333} 2} & {\color[HTML]{333333} 0} & {\color[HTML]{333333} 0.0} & {\color[HTML]{333333} 0} & - \\ \hline
{\color[HTML]{333333} \textbf{Dental decay}} & {\color[HTML]{333333} 24} & {\color[HTML]{333333} 2} & {\color[HTML]{333333} 0.08} & {\color[HTML]{333333} 2} & 1 \\ \hline
{\color[HTML]{333333} \textbf{Depression}} & {\color[HTML]{333333} 82} & {\color[HTML]{333333} 24} & {\color[HTML]{333333} 0.29} & {\color[HTML]{333333} 24} & 1 \\ \hline
{\color[HTML]{333333} \textbf{Diagnosed vitamin D deficiency}} & {\color[HTML]{333333} 4} & {\color[HTML]{333333} 1} & {\color[HTML]{333333} 0.25} & {\color[HTML]{333333} 1} & 1 \\ \hline
\textbf{Diagnosed with sleep apnea} & 6 & 1 & 0.17 & 2 & 0.5 \\ \hline
\textbf{Dyslexia} & 2 & 1 & 0.5 & 1 & 1 \\ \hline
\textbf{Earlobe free or attached} & 11 & 0 & 0.0 & 0 & - \\ \hline
\textbf{Hair type} & 7 & 1 & 0.14 & 0 & - \\ \hline
\textbf{Hypertension} & 52 & 13 & 0.25 & 13 & 1 \\ \hline
\textbf{Hypertriglyceridemia} & 4 & 0 & 0.0 & 0 & - \\ \hline
\textbf{Irritable bowel syndrome} & 2 & 0 & 0.0 & 1 & 0 \\ \hline
\textbf{Mental disease} & 180 & 36 & 0.2 & 45 & 0.8 \\ \hline
\textbf{Migraine} & 25 & 5 & 0.2 & 5 & 1 \\ \hline
\textbf{Motion sickness} & 3 & 0 & 0.0 & 0 & - \\ \hline
\textbf{Panic disorder} & 2 & 0 & 0.0 & 0 & - \\ \hline
\textbf{Photic sneeze reflex (photoptarmic reflex)} & 4 & 0 & 0.0 & 1 & 0 \\ \hline
\textbf{PTSD} & 5 & 0 & 0.0 & 0 & - \\ \hline
\textbf{Scoliosis} & 29 & 1 & 0.03 & 1 & 1 \\ \hline
\textbf{Sensitivity to Mosquito bites} & 3 & 0 & 0.0 & 0 & - \\ \hline
\textbf{Sleep disorders} & 93 & 1 & 0.01 & 1 & 1 \\ \hline
\textbf{Strabismus} & 1 & 0 & 0.0 & 0 & - \\ \hline
\textbf{Thyroid issues cancer} & 8 & 0 & 0.0 & 0 & - \\ \hline
\textbf{TypeIIDiabetes} & 193 & 5 & 0.03 & 10 & 0.5 \\ \hline
\textbf{Eczema} & 15 & 3 & 0.2 & 3 & 1 \\ \hline
\textbf{Restless leg syndrome} & 6 & 2 & 0.33 & 2 & 1 \\ \hline
\end{tabular}
\caption{\textbf{A table showing the new gene identification ratio associated with each phenotype.} In this table, the first column shows the phenotype, the second column \texttt{(A)} shows the number of common genes between our genotype data and the one downloaded from the GWAS Catalog, the third column \texttt{(I)} shows the number of common genes identified by machine/deep learning and the one listed in the second column \texttt{(A)}, and the fourth column \texttt{(GIR)} shows the gene identification ratio for each phenotype as specified in Equation \ref{geneidentificationratio} and calculated by dividing the third column by the second column (\texttt{(I/A)}). The fifth column \texttt{(A2)} shows the number of common genes between the one downloaded from the GWAS Catalog and identified by machine/deep learning after p-values thresholding, and the last column shows the final gene identification ratio calculated by dividing \texttt{(I/A2)}. "-" indicates that the values in A2 are 0, resulting in division by zero.}
\label{newgeneidentificationratio}
\end{table*}

Table \ref{rann} shows the correlation between the New GIR (see Table \ref{newgeneidentificationratio}, column \texttt{New GIR}) and the performance of machine/deep learning algorithms in terms of AUC, F1 Score, and MCC. It suggests that a deep learning algorithm optimized for the F1 Score achieves a higher identification ratio than models optimized for other evaluation metrics.

\begin{table}[!ht]
\begin{tabular}{|l|l|l|}
\hline
\rowcolor[HTML]{FFFFFF} 
\textbf{} & \textbf{Machine learning} & \textbf{Deep learning} \\ \hline
\textbf{AUC} & 0.06 & 0.34 \\ \hline
\textbf{F1 Score} & 0.23 & 0.40 \\ \hline
\textbf{MCC} & 0.24 & 0.35 \\ \hline
\end{tabular}
\caption{A table showing the correlation between the New GIR and the performance of the machine/deep learning algorithms.}
\label{rann}
\end{table}

Supplementary material (Section 3) lists the common genes between the ones identified by the machine and deep learning algorithms and those listed on the GWAS Catalog for each phenotype. The phenotypes, the number of associated genes from the GWAS Catalog for each phenotype, and the number of common genes and SNPs identified by ML/DL and from the GWAS Catalog for each phenotype are available on GitHub (Final\_Results.pdf).

We identified the common SNPs and genes among all phenotypes. Identifying the shared genes between phenotypes is of great importance. It helps identify shared gene influences across multiple phenotypes and genes associated with more than one phenotype.

\begin{figure*}[!ht]
\centering
 \includegraphics[width=13cm,height=10cm]{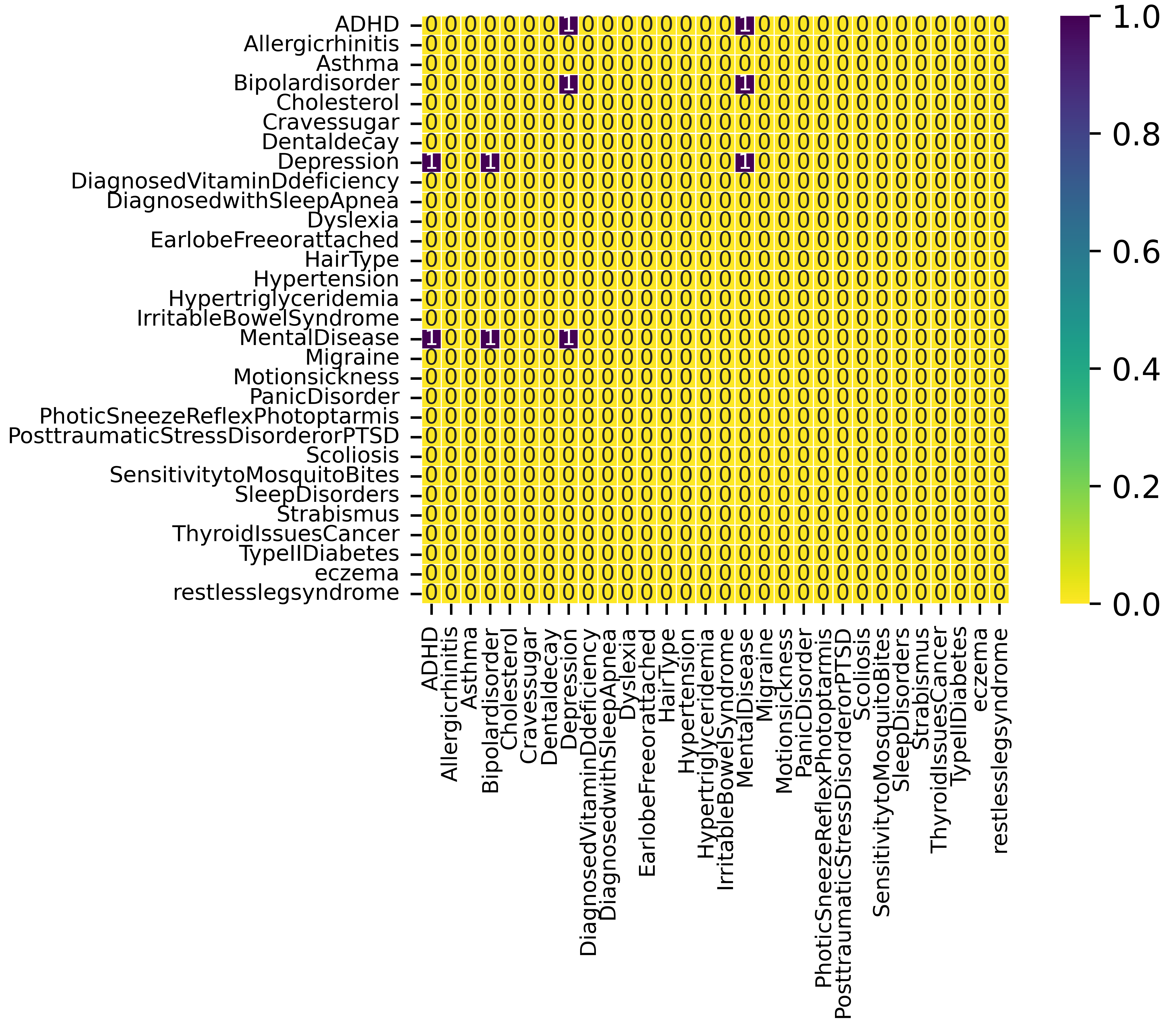}
\caption{A heatmap showing the number of common SNPs between phenotypes.}
\label{intersectionsnps}
\end{figure*}

\begin{figure*}[!ht]
\centering
 \includegraphics[width=13cm,height=10cm]{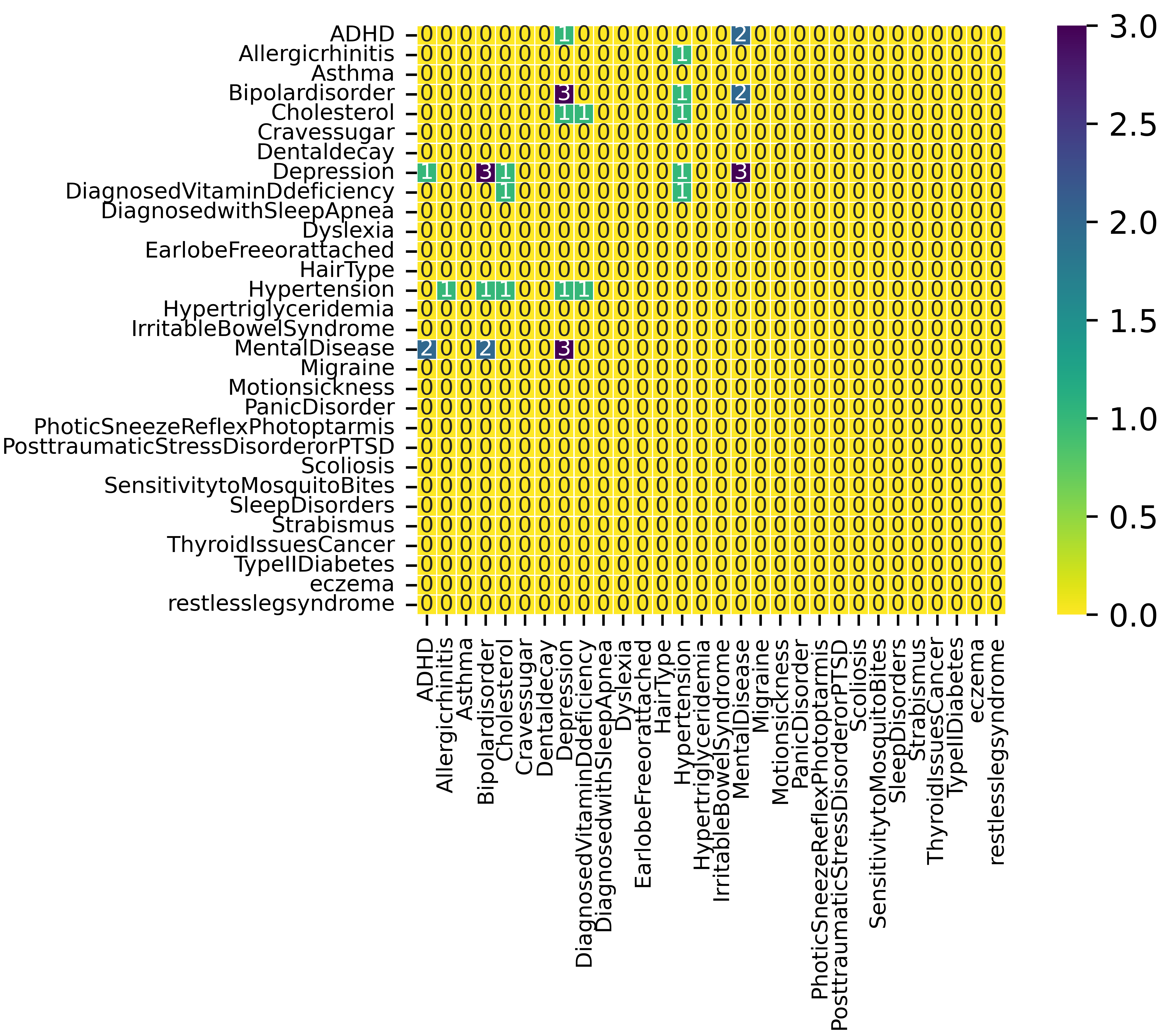}
\caption{A heatmap showing the number of common genes between phenotypes.}
\label{intersectiongenes}
\end{figure*}

We have presented two heatmaps illustrating the overlap of genetic factors among phenotypes: one depicting the number of common SNPs (Figure \ref{intersectionsnps}), and the other showing the number of common genes (Figure \ref{intersectiongenes}). The number of common genes (Figure \ref{intersectiongenes}) between two phenotypes is greater than the number of common SNPs (Figure \ref{intersectionsnps}) because the actual associated SNPs can be different, but the genes in which these SNPs are present can be the same.

Figure \ref{intersectionsnps} shows one common SNP between {Depression, Mental Disease} and ADHD, which is also a neuro-developmental disorder. There is also one common SNP between Depression, Mental Disease, and Bipolar Disorder, a mental health condition. There is a common SNP between Depression and Mental Disease. These findings suggest that machine-learning and deep-learning algorithms can identify risk variants for some phenotypes and even common risk variants across diseases.

Figure \ref{intersectiongenes} shows one common gene between hypertension and {Allergic Rhinitis, Bipolar Disorder, Cholesterol, Depression, and Vitamin D Deficiency}. 


\section{Conclusion}
\label{discussion}

We have proposed a pipeline to identify genes associated with phenotypes using ML and DL algorithms on openSNP data. The identified genes were compared with the existing associated genes from the GWAS Catalog. The mean of the per-phenotype gene identification ratios was 0.84, though this average is sensitive to phenotypes with very few overlapping genes and should be interpreted alongside the individual ratios. The classification performance of the best-performing models was used to assess the reliability of the identified genes.

In the ML/DL approach, combining two or more SNPs, which may be biologically irrelevant, can encode more information than a single SNP selected by GWAS, thereby reducing the gene identification ratio. The quantity and quality of genotype data can affect the genes identified by machine learning, and results can be improved by including all SNPs and genes in the dataset. 

We observed that p-value thresholds affect gene identification performance. The optimal number of SNPs for classification performance is limited. In that case, the SNPs associated with the phenotypes downloaded from the GWAS Catalog may not even appear in the final results. SNPs from the GWAS Catalog for a particular phenotype are collected across multiple studies and data sources, and it is likely that none of these SNPs are significant for the sample dataset. Although other models perform well when a large number of SNPs are used for training, this undermines the study's rationale, which suggests choosing the model that achieves high performance with a minimal number of SNPs.

Our main focus was to identify genes associated with specific phenotypes using machine learning and compare them with existing genes associated with the same phenotype. It is possible that the identified genes differ, but rather than using machine learning for identification, we can use the proposed pipeline as a pre-processing step for GWAS, allowing exploration of various genomic regions that can be further scanned to identify potential association candidates.

Calculating feature weights is time-consuming during training deep learning models, so we recommend first identifying the best-performing algorithms and retraining those models to obtain the weights. 

The code for this study is available on GitHub \url{https://github.com/MuhammadMuneeb007/Identifying-genes-associated-with-30-phenotypes-using-machine-deep-learning-}.

  \begin{mdframed}[linewidth=1pt,linecolor=black,
  innerleftmargin=8pt,innerrightmargin=8pt,
  innertopmargin=16pt-8.2pt,innerbottommargin=6pt]
  {\fontsize{8.2pt}{10pt}\bfseries Key Points\par}
\begin{adjustwidth}{8pt}{0cm}
     \begin{itemize}
 \item This study proposes a pipeline to identify genes associated with phenotypes using machine-learning and deep-learning algorithms.
  \item The best-performing models, optimized for evaluation metrics such as AUC, MCC, and F1-Score, can be combined to identify important features or SNPs that assist in gene identification.
    \item Genotype data quality, the population under study, the number of variants, missingness per person, and the p-value threshold of the best-performing models affect the gene identification ratio.
    \item The mean of the per-phenotype gene identification ratios was 0.84, though individual ratios vary widely across phenotypes.
\end{itemize}
\end{adjustwidth}
\end{mdframed}

\clearpage

\section{Competing interests}
The authors declare that they have no competing interests.

\section{Author contributions statement}
M.M. wrote the first draft of the manuscript. M.M. and Y.M. analyzed the results. D.A., M.M., and Y.M. reviewed and edited the manuscript. M.M. and D.A. contributed to the manuscript's methodology.

\section{Data availability}
The dataset used in this study is available from openSNP (\url{https://opensnp.org/}), and the code is available on GitHub (\url{https://github.com/MuhammadMuneeb007/Identifying-genes-associated-with-30-phenotypes-using-machine-deep-learning-}).

\section{Acknowledgments}
Not applicable.
\bibliographystyle{unsrt}
\bibliography{reference}

@article{Leal2019,
  title = {Identification of disease-associated loci using machine learning for genotype and network data integration},
  volume = {35},
  ISSN = {1367-4811},
  url = {http://dx.doi.org/10.1093/bioinformatics/btz310},
  DOI = {10.1093/bioinformatics/btz310},
  number = {24},
  journal = {Bioinformatics},
  publisher = {Oxford University Press (OUP)},
  author = {Leal,  Luis G and David,  Alessia and Jarvelin,  Marjo-Riita and Sebert,  Sylvain and M\"{a}nnikk\"{o},  Minna and Karhunen,  Ville and Seaby,  Eleanor and Hoggart,  Clive and Sternberg,  Michael J E},
  editor = {Valencia,  Alfonso},
  year = {2019},
  month = may,
  pages = {5182–5190}
}

@article{Wu2010,
  doi = {10.1186/gb-2010-11-5-r53},
  url = {https://doi.org/10.1186/gb-2010-11-5-r53},
  year = {2010},
  publisher = {Springer Science and Business Media {LLC}},
  volume = {11},
  number = {5},
  pages = {R53},
  author = {Guanming Wu and Xin Feng and Lincoln Stein},
  title = {A human functional protein interaction network and its application to cancer data analysis},
  journal = {Genome Biology}
}

@article{Cirillo2017,
  doi = {10.3389/fgene.2017.00174},
  url = {https://doi.org/10.3389/fgene.2017.00174},
  year = {2017},
  month = nov,
  publisher = {Frontiers Media {SA}},
  volume = {8},
  author = {Elisa Cirillo and Laurence D. Parnell and Chris T. Evelo},
  title = {A Review of Pathway-Based Analysis Tools That Visualize Genetic Variants},
  journal = {Frontiers in Genetics}
}

@article{Roberts2022,
  doi = {10.1093/nargab/lqab124},
  url = {https://doi.org/10.1093/nargab/lqab124},
  year = {2022},
  month = jan,
  publisher = {Oxford University Press ({OUP})},
  volume = {4},
  number = {1},
  author = {Aedan G K Roberts and Daniel R Catchpoole and Paul J Kennedy},
  title = {Identification of differentially distributed gene expression and distinct sets of cancer-related genes identified by changes in mean and variability},
  journal = {{NAR} Genomics and Bioinformatics}
}

@article{Alqudah2020,
  doi = {10.1016/j.jare.2019.10.013},
  url = {https://doi.org/10.1016/j.jare.2019.10.013},
  year = {2020},
  month = mar,
  publisher = {Elsevier {BV}},
  volume = {22},
  pages = {119--135},
  author = {Ahmad M. Alqudah and Ahmed Sallam and P. Stephen Baenziger and Andreas B\"{o}rner},
  title = {{GWAS}: Fast-forwarding gene identification and characterization in temperate Cereals: lessons from Barley {\textendash} A review},
  journal = {Journal of Advanced Research}
}

@article{Fickett1996,
  doi = {10.1016/s0097-8485(96)80012-x},
  url = {https://doi.org/10.1016/s0097-8485(96)80012-x},
  year = {1996},
  month = mar,
  publisher = {Elsevier {BV}},
  volume = {20},
  number = {1},
  pages = {103--118},
  author = {James W. Fickett},
  title = {The gene identification problem: An overview for developers},
  journal = {Computers {\&}amp; Chemistry}
}

@article{Antonarakis2021,
  doi = {10.1002/ajmg.a.62400},
  url = {https://doi.org/10.1002/ajmg.a.62400},
  year = {2021},
  month = jun,
  publisher = {Wiley},
  volume = {185},
  number = {11},
  pages = {3266--3275},
  author = {Stylianos E. Antonarakis},
  title = {History of the methodology of disease gene identification},
  journal = {American Journal of Medical Genetics Part A}
}

@incollection{Maran2020,
  doi = {10.4018/978-1-7998-3095-5.ch006},
  url = {https://doi.org/10.4018/978-1-7998-3095-5.ch006},
  year = {2020},
  publisher = {{IGI} Global},
  pages = {129--144},
  author = {Pyingkodi Maran and Shanthi S. and Thenmozhi K. and Hemalatha D. and Nanthini K.},
  title = {A Novel Deep Learning Method for Identification of Cancer Genes From Gene Expression Dataset},
  booktitle = {Machine Learning and Deep Learning in Real-Time Applications}
}

@article{Abbas2020,
  doi = {10.1186/s12920-020-00771-4},
  url = {https://doi.org/10.1186/s12920-020-00771-4},
  year = {2020},
  month = aug,
  publisher = {Springer Science and Business Media {LLC}},
  volume = {13},
  number = {1},
  author = {Mostafa Abbas and Yasser EL-Manzalawy},
  title = {Machine learning based refined differential gene expression analysis of pediatric sepsis},
  journal = {{BMC} Medical Genomics}
}

@article{McKinney2006,
  doi = {10.2165/00822942-200605020-00002},
  url = {https://doi.org/10.2165/00822942-200605020-00002},
  year = {2006},
  publisher = {Springer Science and Business Media {LLC}},
  volume = {5},
  number = {2},
  pages = {77--88},
  author = {Brett A McKinney and David M Reif and Marylyn D Ritchie and Jason H Moore},
  title = {Machine Learning for Detecting Gene-Gene Interactions},
  journal = {Applied Bioinformatics}
}

@article{Mahood2020,
  doi = {10.1002/aps3.11376},
  url = {https://doi.org/10.1002/aps3.11376},
  year = {2020},
  month = jul,
  publisher = {Wiley},
  volume = {8},
  number = {7},
  author = {Elizabeth H. Mahood and Lars H. Kruse and Gaurav D. Moghe},
  title = {Machine learning: A powerful tool for gene function prediction in plants},
  journal = {Applications in Plant Sciences}
}

@INPROCEEDINGS{9802470,  author={Muneeb, Muhammad and Feng, Samuel F. and Henschel, Andreas},  booktitle={2022 10th International Conference on Bioinformatics and Computational Biology (ICBCB)},   title={Tutorial on 8 Genotype Files Conversion},   year={2022},  volume={},  number={},  pages={13-17},  doi={10.1109/ICBCB55259.2022.9802470}}

@article{Greshake2014,
  doi = {10.1371/journal.pone.0089204},
  url = {https://doi.org/10.1371/journal.pone.0089204},
  year = {2014},
  month = mar,
  publisher = {Public Library of Science ({PLoS})},
  volume = {9},
  number = {3},
  pages = {e89204},
  author = {Bastian Greshake and Philipp E. Bayer and Helge Rausch and Julia Reda},
  editor = {Tricia A. Thornton-Wells},
  title = {{openSNP}{\textendash}A Crowdsourced Web Resource for Personal Genomics},
  journal = {{PLoS} {ONE}}
}

@article{scikit-learn,
 title={Scikit-learn: Machine Learning in {P}ython},
 author={Pedregosa, F. and Varoquaux, G. and Gramfort, A. and Michel, V.
         and Thirion, B. and Grisel, O. and Blondel, M. and Prettenhofer, P.
         and Weiss, R. and Dubourg, V. and Vanderplas, J. and Passos, A. and
         Cournapeau, D. and Brucher, M. and Perrot, M. and Duchesnay, E.},
 journal={Journal of Machine Learning Research},
 volume={12},
 pages={2825--2830},
 year={2011}
}

@incollection{Cristianini2008,
  doi = {10.1007/978-0-387-30162-4_415},
  url = {https://doi.org/10.1007/978-0-387-30162-4_415},
  year = {2008},
  publisher = {Springer {US}},
  pages = {928--932},
  author = {Nello Cristianini and Elisa Ricci},
  title = {Support Vector Machines},
  booktitle = {Encyclopedia of Algorithms}
}

@article{asd,
author = {Jijo, Bahzad and Mohsin Abdulazeez, Adnan},
year = {2021},
month = {01},
pages = {20-28},
title = {Classification Based on Decision Tree Algorithm for Machine Learning},
volume = {2},
journal = {Journal of Applied Science and Technology Trends}
}

@inproceedings{Chen:2016:XST:2939672.2939785,
 author = {Chen, Tianqi and Guestrin, Carlos},
 title = {{XGBoost}: A Scalable Tree Boosting System},
 booktitle = {Proceedings of the 22nd ACM SIGKDD International Conference on Knowledge Discovery and Data Mining},
 series = {KDD '16},
 year = {2016},
 isbn = {978-1-4503-4232-2},
 location = {San Francisco, California, USA},
 pages = {785--794},
 numpages = {10},
 url = {http://doi.acm.org/10.1145/2939672.2939785},
 doi = {10.1145/2939672.2939785},
 acmid = {2939785},
 publisher = {ACM},
 address = {New York, NY, USA},
}

@article{mcculloch1943logical,
  title={A logical calculus of the ideas immanent in nervous activity},
  author={McCulloch, Warren S and Pitts, Walter},
  journal={The bulletin of mathematical biophysics},
  volume={5},
  number={4},
  pages={115--133},
  year={1943},
  publisher={Springer}
}

@INPROCEEDINGS{8053243,
  author={Dey, Rahul and Salem, Fathi M.},
  booktitle={2017 IEEE 60th International Midwest Symposium on Circuits and Systems (MWSCAS)}, 
  title={Gate-variants of Gated Recurrent Unit (GRU) neural networks}, 
  year={2017},
  volume={},
  number={},
  pages={1597-1600},
  doi={10.1109/MWSCAS.2017.8053243}}

@article{Hochreiter1997,
  doi = {10.1162/neco.1997.9.8.1735},
  url = {https://doi.org/10.1162/neco.1997.9.8.1735},
  year = {1997},
  month = nov,
  publisher = {{MIT} Press - Journals},
  volume = {9},
  number = {8},
  pages = {1735--1780},
  author = {Sepp Hochreiter and J\"{u}rgen Schmidhuber},
  title = {Long Short-Term Memory},
  journal = {Neural Computation}
}

@article{Schuster1997,
  doi = {10.1109/78.650093},
  url = {https://doi.org/10.1109/78.650093},
  year = {1997},
  publisher = {Institute of Electrical and Electronics Engineers ({IEEE})},
  volume = {45},
  number = {11},
  pages = {2673--2681},
  author = {M. Schuster and K.K. Paliwal},
  title = {Bidirectional recurrent neural networks},
  journal = {{IEEE} Transactions on Signal Processing}
}

@article{Mori2021,
  doi = {10.1038/s41598-021-95969-6},
  url = {https://doi.org/10.1038/s41598-021-95969-6},
  year = {2021},
  month = aug,
  publisher = {Springer Science and Business Media {LLC}},
  volume = {11},
  number = {1},
  author = {Yasukuni Mori and Hajime Yokota and Isamu Hoshino and Yosuke Iwatate and Kohei Wakamatsu and Takashi Uno and Hiroki Suyari},
  title = {Deep learning-based gene selection in comprehensive gene analysis in pancreatic cancer},
  journal = {Scientific Reports}
}

@article{Mi2021,
  doi = {10.1038/s41467-021-22756-2},
  url = {https://doi.org/10.1038/s41467-021-22756-2},
  year = {2021},
  month = may,
  publisher = {Springer Science and Business Media {LLC}},
  volume = {12},
  number = {1},
  author = {Xinlei Mi and Baiming Zou and Fei Zou and Jianhua Hu},
  title = {Permutation-based identification of important biomarkers for complex diseases via machine learning models},
  journal = {Nature Communications}
}

@article{Musolf2021,
  doi = {10.1007/s00439-021-02402-z},
  url = {https://doi.org/10.1007/s00439-021-02402-z},
  year = {2021},
  month = dec,
  publisher = {Springer Science and Business Media {LLC}},
  volume = {141},
  number = {9},
  pages = {1515--1528},
  author = {Anthony M. Musolf and Emily R. Holzinger and James D. Malley and Joan E. Bailey-Wilson},
  title = {What makes a good prediction? Feature importance and beginning to open the black box of machine learning in genetics},
  journal = {Human Genetics}
}

@article{Pudjihartono2022,
  doi = {10.3389/fbinf.2022.927312},
  url = {https://doi.org/10.3389/fbinf.2022.927312},
  year = {2022},
  month = jun,
  publisher = {Frontiers Media {SA}},
  volume = {2},
  author = {Nicholas Pudjihartono and Tayaza Fadason and Andreas W. Kempa-Liehr and Justin M. O{\textquotesingle}Sullivan},
  title = {A Review of Feature Selection Methods for Machine Learning-Based Disease Risk Prediction},
  journal = {Frontiers in Bioinformatics}
}

@article{Alatrany2023,
  doi = {10.1371/journal.pone.0283712},
  url = {https://doi.org/10.1371/journal.pone.0283712},
  year = {2023},
  month = may,
  publisher = {Public Library of Science ({PLoS})},
  volume = {18},
  number = {5},
  pages = {e0283712},
  author = {Abbas Saad Alatrany and Wasiq Khan and Abir Hussain and Dhiya Al-Jumeily and},
  editor = {Mohamed Hammad},
  title = {Wide and deep learning based approaches for classification of Alzheimer's disease using genome-wide association studies},
  journal = {{PLOS} {ONE}}
}

@article{Gomes2022,
  doi = {10.3390/genes13091557},
  url = {https://doi.org/10.3390/genes13091557},
  year = {2022},
  month = aug,
  publisher = {{MDPI} {AG}},
  volume = {13},
  number = {9},
  pages = {1557},
  author = {Rahul Gomes and Nijhum Paul and Nichol He and Aaron Francis Huber and Rick J. Jansen},
  title = {Application of Feature Selection and Deep Learning for Cancer Prediction Using {DNA} Methylation Markers},
  journal = {Genes}
}

@article{Tabl2019,
  doi = {10.3389/fgene.2019.00256},
  url = {https://doi.org/10.3389/fgene.2019.00256},
  year = {2019},
  month = mar,
  publisher = {Frontiers Media {SA}},
  volume = {10},
  author = {Ashraf Abou Tabl and Abedalrhman Alkhateeb and Waguih ElMaraghy and Luis Rueda and Alioune Ngom},
  title = {A Machine Learning Approach for Identifying Gene Biomarkers Guiding the Treatment of Breast Cancer},
  journal = {Frontiers in Genetics}
}

@article{Mahendran2020,
  doi = {10.3389/fgene.2020.603808},
  url = {https://doi.org/10.3389/fgene.2020.603808},
  year = {2020},
  month = dec,
  publisher = {Frontiers Media {SA}},
  volume = {11},
  author = {Nivedhitha Mahendran and P. M. Durai Raj Vincent and Kathiravan Srinivasan and Chuan-Yu Chang},
  title = {Machine Learning Based Computational Gene Selection Models: A Survey,  Performance Evaluation,  Open Issues,  and Future Research Directions},
  journal = {Frontiers in Genetics}
}

@article{Andrews2023,
  doi = {10.1016/j.ebiom.2023.104511},
  url = {https://doi.org/10.1016/j.ebiom.2023.104511},
  year = {2023},
  month = apr,
  publisher = {Elsevier {BV}},
  volume = {90},
  pages = {104511},
  author = {Shea J. Andrews and Alan E. Renton and Brian Fulton-Howard and Anna Podlesny-Drabiniok and Edoardo Marcora and Alison M. Goate},
  title = {The complex genetic architecture of Alzheimer{\textquotesingle}s disease: novel insights and future directions},
  journal = {{eBioMedicine}}
}

@article{Kia2021,
  doi = {10.1001/jamaneurol.2020.5257},
  url = {https://doi.org/10.1001/jamaneurol.2020.5257},
  year = {2021},
  month = apr,
  publisher = {American Medical Association ({AMA})},
  volume = {78},
  number = {4},
  pages = {464},
  author = {Demis A. Kia and David Zhang and Sebastian Guelfi and Claudia Manzoni and Leon Hubbard and Regina H. Reynolds and Juan Bot{\'{\i}}a and Mina Ryten and Raffaele Ferrari and Patrick A. Lewis and Nigel Williams and Daniah Trabzuni and John Hardy and Nicholas W. Wood and Alastair J. Noyce and Rauan Kaiyrzhanov and Ben Middlehurst and Demis A. Kia and Manuela Tan and Henry Houlden and Huw R. Morris and Helene Plun-Favreau and Peter Holmans and John Hardy and Daniah Trabzuni and Jose Bras and John Quinn PhD and Kin Y. Mok and Kerri J. Kinghorn and Kimberley Billingsley and Nicholas W. Wood and Patrick Lewis and Sebastian Schreglmann and Rita Guerreiro and Ruth Lovering and Lea R{\textquotesingle}Bibo and Claudia Manzoni and Mie Rizig and Mina Ryten and Sebastian Guelfi and Valentina Escott-Price and Viorica Chelban and Thomas Foltynie and Nigel Williams and Alexis Brice and Fabrice Danjou and Suzanne Lesage and Jean-Christophe Corvol and Maria Martinez and Claudia Schulte and Kathrin Brockmann and Javier Sim{\'{o}}n-S{\'{a}}nchez and Peter Heutink and Patrizia Rizzu and Manu Sharma and Thomas Gasser and Aude Nicolas and Mark R. Cookson and Sara Bandres-Ciga and Cornelis Blauwendraat and David W. Craig and Faraz Faghri and J. Raphael Gibbs and Dena G. Hernandez and Kendall Van Keuren-Jensen and Joshua M. Shulman and Hampton L. Leonard and Mike A. Nalls and Laurie Robak and Steven Lubbe and Steven Finkbeiner and Niccolo E. Mencacci and Codrin Lungu and Andrew B Singleton and Sonja W. Scholz and Xylena Reed and Roy N. Alcalay and Ziv Gan-Or and Guy A. Rouleau and Lynne Krohn and Jacobus J. van Hilten and Johan Marinus and Astrid D. Adarmes-G{\'{o}}mez and Miquel Aguilar and Ignacio Alvarez and Victoria Alvarez and Francisco Javier Barrero and Jes{\'{u}}s A. Bergareche Yarza and Inmaculada Bernal-Bernal and Marta Blazquez and Marta Bonilla-Toribio and Juan A. Bot{\'{\i}}a and Mar{\'{\i}}a T. Boungiorno and Dolores Buiza-Rueda and Ana C{\`{a}}mara and F{\'{a}}tima Carrillo and Mario Carri{\'{o}}n-Claro and Debora Cerdan and Jordi Clarim{\'{o}}n and Yaroslau Compta and Monica Diez-Fairen and Oriol Dols-Icardo and Jacinto Duarte and Raquel Duran and Francisco Escamilla-Sevilla and Mario Ezquerra and Cici Feliz and Manel Fern{\`{a}}ndez and Rub{\'{e}}n Fern{\`{a}}ndez-Santiago and Ciara Garcia and Pedro Garc{\'{\i}}a-Ruiz and Pilar G{\'{o}}mez-Garre and Maria J. Gomez Heredia and Isabel Gonzalez-Aramburu and Ana G. Pagola and Janet Hoenicka and Jon Infante and Adriano Jimenez-Escrig and Jaime Kulisevsky and Miguel A. Labrador-Espinosa and Jose Luis Lopez-Sendon and Adolfo L{\'{o}}pez de Munain Arregui and Daniel Macias and Irene Mart{\'{\i}}nez Torres and Juan Mar{\'{\i}}n and Maria Jose Marti and Juan Carlos Mart{\'{\i}}nez-Castrillo and Carlota M{\`{e}}ndez-del-Barrio and Manuel Men{\'{e}}ndez Gonz{\'{a}}lez and Marina Mata Adolfo M{\'{\i}}nguez and Pablo Mir and Elisabet Mondragon Rezola and Esteban Mu{\~{n}}oz and Javier Pagonabarraga and Pau Pastor and Francisco Perez Errazquin and Teresa Perin{\'{a}}n-Tocino and Javier Ruiz-Mart{\'{\i}}nez and Clara Ruz and Antonio Sanchez Rodriguez and Mar{\'{\i}}a Sierra and Esther Suarez-Sanmartin and Cesar Tabernero and Juan Pablo Tartari and Cristina Tejera-Parrado and Eduard Tolosa and Francesc Valldeoriola and Laura Vargas-Gonz{\'{a}}lez and Lydia Vela and Francisco Vives and Alexander Zimprich and Lasse Pihlstrom and Mathias Toft and Sulev Koks and Pille Taba and Sharon Hassin-Baer and Michael Weale and Adaikalavan Ramasamy and Colin Smith and Manuel Sebastian Guelfi and Karishma D{\textquotesingle}sa and Paola Forabosco and Juan A. Boti{\'{a}} and},
  title = {Identification of Candidate Parkinson Disease Genes by Integrating Genome-Wide Association Study,  Expression,  and Epigenetic Data Sets},
  journal = {{JAMA} Neurology}
}

@article{Udosen2023,
  doi = {10.3390/ijms24032164},
  url = {https://doi.org/10.3390/ijms24032164},
  year = {2023},
  month = jan,
  publisher = {{MDPI} {AG}},
  volume = {24},
  number = {3},
  pages = {2164},
  author = {Brenda Udosen and Opeyemi Soremekun and Abram Kamiza and Tafadzwa Machipisa and Cisse Cheickna and Olaposi Omotuyi and Mahmoud Soliman and Mamadou W{\'{e}}l{\'{e}} and Oyekanmi Nashiru and Tinashe Chikowore and Segun Fatumo},
  title = {Meta-Analysis and Multivariate {GWAS} Analyses in 77, 850 Individuals of African Ancestry Identify Novel Variants Associated with Blood Pressure Traits},
  journal = {International Journal of Molecular Sciences}
}

@article{CanoGamez2020,
  doi = {10.3389/fgene.2020.00424},
  url = {https://doi.org/10.3389/fgene.2020.00424},
  year = {2020},
  month = may,
  publisher = {Frontiers Media {SA}},
  volume = {11},
  author = {Eddie Cano-Gamez and Gosia Trynka},
  title = {From {GWAS} to Function: Using Functional Genomics to Identify the Mechanisms Underlying Complex Diseases},
  journal = {Frontiers in Genetics}
}

@article{Li2021,
  doi = {10.3389/fgene.2021.713230},
  url = {https://doi.org/10.3389/fgene.2021.713230},
  year = {2021},
  month = sep,
  publisher = {Frontiers Media {SA}},
  volume = {12},
  author = {Binglan Li and Marylyn D. Ritchie},
  title = {From {GWAS} to Gene: Transcriptome-Wide Association Studies and Other Methods to Functionally Understand {GWAS} Discoveries},
  journal = {Frontiers in Genetics}
}

@article{Uygun2016,
  doi = {10.1371/journal.pcbi.1005244},
  url = {https://doi.org/10.1371/journal.pcbi.1005244},
  year = {2016},
  month = dec,
  publisher = {Public Library of Science ({PLoS})},
  volume = {12},
  number = {12},
  pages = {e1005244},
  author = {Sahra Uygun and Cheng Peng and Melissa D. Lehti-Shiu and Robert L. Last and Shin-Han Shiu},
  editor = {Xianghong Jasmine Zhou},
  title = {Utility and Limitations of Using Gene Expression Data to Identify Functional Associations},
  journal = {{PLOS} Computational Biology}
}

@article{Xu2020,
  doi = {10.1155/2020/8082697},
  url = {https://doi.org/10.1155/2020/8082697},
  year = {2020},
  month = may,
  publisher = {Hindawi Limited},
  volume = {2020},
  pages = {1--13},
  author = {Houxi Xu and Yuzhu Ma and Jinzhi Zhang and Jialin Gu and Xinyue Jing and Shengfeng Lu and Shuping Fu and Jiege Huo},
  title = {Identification and Verification of Core Genes in Colorectal Cancer},
  journal = {{BioMed} Research International}
}

@article{Lu2018,
  doi = {10.1111/and.13169},
  url = {https://doi.org/10.1111/and.13169},
  year = {2018},
  month = oct,
  publisher = {Hindawi Limited},
  volume = {51},
  number = {1},
  pages = {e13169},
  author = {Wenzong Lu and Zhe Ding},
  title = {Identification of key genes in prostate cancer gene expression profile by bioinformatics},
  journal = {Andrologia}
}

@incollection{Ghadle2020,
  doi = {10.1007/978-981-15-7421-4_23},
  url = {https://doi.org/10.1007/978-981-15-7421-4\_23},
  year = {2020},
  month = oct,
  publisher = {Springer Singapore},
  pages = {253--259},
  author = {Saroj Ghadle and Rakesh Tripathi and Sanjay Kumar and Vandana Munde},
  title = {Study on Analysis of Gene Expression Dataset and Identification of Differentially Expressed Genes},
  booktitle = {Intelligent Computing and Networking}
}

@article{Begum2021,
  doi = {10.1016/j.eswa.2021.114914},
  url = {https://doi.org/10.1016/j.eswa.2021.114914},
  year = {2021},
  month = sep,
  publisher = {Elsevier {BV}},
  volume = {177},
  pages = {114914},
  author = {Shemim Begum and Ram Sarkar and Debasis Chakraborty and Sagnik Sen and Ujjwal Maulik},
  title = {Application of active learning in {DNA} microarray data for cancerous gene identification},
  journal = {Expert Systems with Applications}
}

@inproceedings{Senadheera2020,
  doi = {10.1109/icter51097.2020.9325446},
  url = {https://doi.org/10.1109/icter51097.2020.9325446},
  year = {2020},
  month = nov,
  publisher = {{IEEE}},
  author = {S. P. B. M Senadheera and A. R. Weerasinghe},
  title = {Hub Genes Identification in Brain Cancer with Gene Expression Data},
  booktitle = {2020 20th International Conference on Advances in {ICT} for Emerging Regions ({ICTer})}
}

@article{Kori2022,
  doi = {10.3390/genes13122233},
  url = {https://doi.org/10.3390/genes13122233},
  year = {2022},
  month = nov,
  publisher = {{MDPI} {AG}},
  volume = {13},
  number = {12},
  pages = {2233},
  author = {Medi Kori and Esra Gov},
  title = {Bioinformatics Prediction and Machine Learning on Gene Expression Data Identifies Novel Gene Candidates in Gastric Cancer},
  journal = {Genes}
}

@article{Anjum2016,
  doi = {10.1089/cmb.2015.0205},
  url = {https://doi.org/10.1089/cmb.2015.0205},
  year = {2016},
  month = apr,
  publisher = {Mary Ann Liebert Inc},
  volume = {23},
  number = {4},
  pages = {239--247},
  author = {Arfa Anjum and Seema Jaggi and Eldho Varghese and Shwetank Lall and Arpan Bhowmik and Anil Rai},
  title = {Identification of Differentially Expressed Genes in RNA-seq Data of Arabidopsis thaliana: A Compound Distribution Approach},
  journal = {Journal of Computational Biology}
}

@incollection{SanSegundoVal2016,
  doi = {10.1007/978-1-4939-3652-6_3},
  url = {https://doi.org/10.1007/978-1-4939-3652-6_3},
  year = {2016},
  publisher = {Springer New York},
  pages = {29--43},
  author = {Ignacio San Segundo-Val and Catalina S. Sanz-Lozano},
  title = {Introduction to the Gene Expression Analysis},
  booktitle = {Methods in Molecular Biology}
}

@article{Gill2022,
  doi = {10.1186/s12870-022-03559-z},
  url = {https://doi.org/10.1186/s12870-022-03559-z},
  year = {2022},
  month = apr,
  publisher = {Springer Science and Business Media {LLC}},
  volume = {22},
  number = {1},
  author = {Mitchell Gill and Robyn Anderson and Haifei Hu and Mohammed Bennamoun and Jakob Petereit and Babu Valliyodan and Henry T. Nguyen and Jacqueline Batley and Philipp E. Bayer and David Edwards},
  title = {Machine learning models outperform deep learning models,  provide interpretation and facilitate feature selection for soybean trait prediction},
  journal = {{BMC} Plant Biology}
}

@inproceedings{NEURIPS2020,
 author = {Wojtas, Maksymilian and Chen, Ke},
 booktitle = {Advances in Neural Information Processing Systems},
 editor = {H. Larochelle and M. Ranzato and R. Hadsell and M.F. Balcan and H. Lin},
 pages = {5105--5114},
 publisher = {Curran Associates, Inc.},
 title = {Feature Importance Ranking for Deep Learning},
 url = {https://proceedings.neurips.cc/paper_files/paper/2020/file/36ac8e558ac7690b6f44e2cb5ef93322-Paper.pdf},
 volume = {33},
 year = {2020}
}

@article{cai2018feature,
  title={Feature selection in machine learning: A new perspective},
  author={Cai, Jie and Luo, Jiawei and Wang, Shulin and Yang, Sheng},
  journal={Neurocomputing},
  volume={300},
  pages={70--79},
  year={2018},
  publisher={Elsevier}
}

@article{FigueroaBarraza2021,
  doi = {10.3390/s21175888},
  url = {https://doi.org/10.3390/s21175888},
  year = {2021},
  month = sep,
  publisher = {{MDPI} {AG}},
  volume = {21},
  number = {17},
  pages = {5888},
  author = {Joaqu{\'{\i}}n Figueroa Barraza and Enrique L{\'{o}}pez Droguett and Marcelo Ramos Martins},
  title = {Towards Interpretable Deep Learning: A Feature Selection Framework for Prognostics and Health Management Using Deep Neural Networks},
  journal = {Sensors}
}

@article{Chen2020,
  doi = {10.1186/s40537-020-00327-4},
  url = {https://doi.org/10.1186/s40537-020-00327-4},
  year = {2020},
  month = jul,
  publisher = {Springer Science and Business Media {LLC}},
  volume = {7},
  number = {1},
  author = {Rung-Ching Chen and Christine Dewi and Su-Wen Huang and Rezzy Eko Caraka},
  title = {Selecting critical features for data classification based on machine learning methods},
  journal = {Journal of Big Data}
}

@article{Saarela2021,
  doi = {10.1007/s42452-021-04148-9},
  url = {https://doi.org/10.1007/s42452-021-04148-9},
  year = {2021},
  month = feb,
  publisher = {Springer Science and Business Media {LLC}},
  volume = {3},
  number = {2},
  author = {Mirka Saarela and Susanne Jauhiainen},
  title = {Comparison of feature importance measures as explanations for classification models},
  journal = {{SN} Applied Sciences}
}

@article{Hicks2022,
  doi = {10.1038/s41598-022-09954-8},
  url = {https://doi.org/10.1038/s41598-022-09954-8},
  year = {2022},
  month = apr,
  publisher = {Springer Science and Business Media {LLC}},
  volume = {12},
  number = {1},
  author = {Steven A. Hicks and Inga Str\"{u}mke and Vajira Thambawita and Malek Hammou and Michael A. Riegler and P{\aa}l Halvorsen and Sravanthi Parasa},
  title = {On evaluation metrics for medical applications of artificial intelligence},
  journal = {Scientific Reports}
}

@article{Gaudillo2019,
  doi = {10.1371/journal.pone.0225574},
  url = {https://doi.org/10.1371/journal.pone.0225574},
  year = {2019},
  month = dec,
  publisher = {Public Library of Science ({PLoS})},
  volume = {14},
  number = {12},
  pages = {e0225574},
  author = {Joverlyn Gaudillo and Jae Joseph Russell Rodriguez and Allen Nazareno and Lei Rigi Baltazar and Julianne Vilela and Rommel Bulalacao and Mario Domingo and Jason Albia},
  editor = {Enrique Hernandez-Lemus},
  title = {Machine learning approach to single nucleotide polymorphism-based asthma prediction},
  journal = {{PLOS} {ONE}}
}

@article{Li2018,
  doi = {10.3389/fgene.2018.00237},
  url = {https://doi.org/10.3389/fgene.2018.00237},
  year = {2018},
  month = jul,
  publisher = {Frontiers Media {SA}},
  volume = {9},
  author = {Bo Li and Nanxi Zhang and You-Gan Wang and Andrew W. George and Antonio Reverter and Yutao Li},
  title = {Genomic Prediction of Breeding Values Using a Subset of {SNPs} Identified by Three Machine Learning Methods},
  journal = {Frontiers in Genetics}
}

@article{Liao2022,
  doi = {10.1145/3506695},
  url = {https://doi.org/10.1145/3506695},
  year = {2022},
  month = apr,
  publisher = {Association for Computing Machinery ({ACM})},
  volume = {31},
  number = {3},
  pages = {1--40},
  author = {Lizhi Liao and Heng Li and Weiyi Shang and Lei Ma},
  title = {An Empirical Study of the Impact of Hyperparameter Tuning and Model Optimization on the Performance Properties of Deep Neural Networks},
  journal = {{ACM} Transactions on Software Engineering and Methodology}
}

@article{Hunter2005,
  doi = {10.1038/nrg1578},
  url = {https://doi.org/10.1038/nrg1578},
  year = {2005},
  month = apr,
  publisher = {Springer Science and Business Media {LLC}},
  volume = {6},
  number = {4},
  pages = {287--298},
  author = {David J. Hunter},
  title = {Gene{\textendash}environment interactions in human diseases},
  journal = {Nature Reviews Genetics}
}

@article{McGue1998,
  doi = {10.1146/annurev.neuro.21.1.1},
  url = {https://doi.org/10.1146/annurev.neuro.21.1.1},
  year = {1998},
  month = mar,
  publisher = {Annual Reviews},
  volume = {21},
  number = {1},
  pages = {1--24},
  author = {Matt McGue and Thomas J. Bouchard},
  title = {{GENETIC} {AND} {ENVIRONMENTAL} {INFLUENCES} {ON} {HUMAN} {BEHAVIORAL} {DIFFERENCES}},
  journal = {Annual Review of Neuroscience}
}

@article{Jelenkovic2020,
  doi = {10.1038/s41598-020-64883-8},
  url = {https://doi.org/10.1038/s41598-020-64883-8},
  year = {2020},
  month = may,
  publisher = {Springer Science and Business Media {LLC}},
  volume = {10},
  number = {1},
  author = {Aline Jelenkovic and Reijo Sund and Yoshie Yokoyama and Antti Latvala and Masumi Sugawara and Mami Tanaka and Satoko Matsumoto and Duarte L. Freitas and Jos{\'{e}} Antonio Maia and Ariel Knafo-Noam and David Mankuta and Lior Abramson and Fuling Ji and Feng Ning and Zengchang Pang and Esther Rebato and Kimberly J. Saudino and Tessa L. Cutler and John L. Hopper and Vilhelmina Ullemar and Catarina Almqvist and Patrik K. E. Magnusson and Wendy Cozen and Amie E. Hwang and Thomas M. Mack and Tracy L. Nelson and Keith E. Whitfield and Joohon Sung and Jina Kim and Jooyeon Lee and Sooji Lee and Clare H. Llewellyn and Abigail Fisher and Emanuela Medda and Lorenza Nistic{\`{o}} and Virgilia Toccaceli and Laura A. Baker and Catherine Tuvblad and Robin P. Corley and Brooke M. Huibregtse and Catherine A. Derom and Robert F. Vlietinck and Ruth J. F. Loos and S. Alexandra Burt and Kelly L. Klump and Judy L. Silberg and Hermine H. Maes and Robert F. Krueger and Matt McGue and Shandell Pahlen and Margaret Gatz and David A. Butler and Jennifer R. Harris and Ingunn Brandt and Thomas S. Nilsen and K. Paige Harden and Elliot M. Tucker-Drob and Carol E. Franz and William S. Kremen and Michael J. Lyons and Paul Lichtenstein and Meike Bartels and Catharina E. M. van Beijsterveldt and Gonneke Willemsen and Sevgi Y. \"{O}ncel and Fazil Aliev and Hoe-Uk Jeong and Yoon-Mi Hur and Eric Turkheimer and Dorret I. Boomsma and Thorkild I. A. S{\o}rensen and Jaakko Kaprio and Karri Silventoinen},
  title = {Genetic and environmental influences on human height from infancy through adulthood at different levels of parental education},
  journal = {Scientific Reports}
}

@article{Marees2018,
  doi = {10.1002/mpr.1608},
  url = {https://doi.org/10.1002/mpr.1608},
  year = {2018},
  month = feb,
  publisher = {Wiley},
  volume = {27},
  number = {2},
  pages = {e1608},
  author = {Andries T. Marees and Hilde de Kluiver and Sven Stringer and Florence Vorspan and Emmanuel Curis and Cynthia Marie-Claire and Eske M. Derks},
  title = {A tutorial on conducting genome-wide association studies: Quality control and statistical analysis},
  journal = {International Journal of Methods in Psychiatric Research}
}

@article{Uffelmann2021,
  doi = {10.1038/s43586-021-00056-9},
  url = {https://doi.org/10.1038/s43586-021-00056-9},
  year = {2021},
  month = aug,
  publisher = {Springer Science and Business Media {LLC}},
  volume = {1},
  number = {1},
  author = {Emil Uffelmann and Qin Qin Huang and Nchangwi Syntia Munung and Jantina de Vries and Yukinori Okada and Alicia R. Martin and Hilary C. Martin and Tuuli Lappalainen and Danielle Posthuma},
  title = {Genome-wide association studies},
  journal = {Nature Reviews Methods Primers}
}

@article{Roy2021,
  doi = {10.1007/s12045-021-1194-0},
  url = {https://doi.org/10.1007/s12045-021-1194-0},
  year = {2021},
  month = jul,
  publisher = {Springer Science and Business Media {LLC}},
  volume = {26},
  number = {7},
  pages = {953--970},
  author = {Sayan Roy and Debanjan Rana},
  title = {Machine Learning in Nonlinear Dynamical Systems},
  journal = {Resonance}
}

@article{McCarthy2008,
  doi = {10.1038/nrg2344},
  year = {2008},
  month = may,
  publisher = {Springer Science and Business Media {LLC}},
  volume = {9},
  number = {5},
  pages = {356--369},
  author = {Mark I. McCarthy and Gon{\c{c}}alo R. Abecasis and Lon R. Cardon and David B. Goldstein and Julian Little and John P. A. Ioannidis and Joel N. Hirschhorn},
  title = {Genome-wide association studies for complex traits: consensus,  uncertainty and challenges},
  journal = {Nature Reviews Genetics}
}

@article{Clayton2005,
  doi = {10.1038/ng1653},
  year = {2005},
  month = oct,
  publisher = {Springer Science and Business Media {LLC}},
  volume = {37},
  number = {11},
  pages = {1243--1246},
  author = {David G Clayton and Neil M Walker and Deborah J Smyth and Rebecca Pask and Jason D Cooper and Lisa M Maier and Luc J Smink and Alex C Lam and Nigel R Ovington and Helen E Stevens and Sarah Nutland and Joanna M M Howson and Malek Faham and Martin Moorhead and Hywel B Jones and Matthew Falkowski and Paul Hardenbol and Thomas D Willis and John A Todd},
  title = {Population structure,  differential bias and genomic control in a large-scale,  case-control association study},
  journal = {Nature Genetics}
}

@article{Liu2019,
  doi = {10.3389/fgene.2019.01091},
  year = {2019},
  month = nov,
  publisher = {Frontiers Media {SA}},
  volume = {10},
  author = {Yang Liu and Duolin Wang and Fei He and Juexin Wang and Trupti Joshi and Dong Xu},
  title = {Phenotype Prediction and Genome-Wide Association Study Using Deep Convolutional Neural Network of Soybean},
  journal = {Frontiers in Genetics}
}

@article{Barreiro2008,
  doi = {10.1038/ng.78},
  url = {https://doi.org/10.1038/ng.78},
  year = {2008},
  month = feb,
  publisher = {Springer Science and Business Media {LLC}},
  volume = {40},
  number = {3},
  pages = {340--345},
  author = {Luis B Barreiro and Guillaume Laval and H{\'{e}}l{\`{e}}ne Quach and Etienne Patin and Llu{\'{\i}}s Quintana-Murci},
  title = {Natural selection has driven population differentiation in modern humans},
  journal = {Nature Genetics}
}

@article{Jaffee_2007,
	doi = {10.1038/sj.mp.4001950},
	url = {https://doi.org/10.1038\%2Fsj.mp.4001950},
	year = 2007,
	month = {jan},
	publisher = {Springer Science and Business Media {LLC}},
	volume = {12},
	number = {5},
	pages = {432--442},
	author = {S R Jaffee and T S Price},
	title = {Gene{\textendash}environment correlations: a review of the evidence and implications for prevention of mental illness},
	journal = {Molecular Psychiatry}
}

@article{Williams_2000,
	doi = {10.1046/j.1523-1755.2000.00982.x},
	url = {https://doi.org/10.1046\%2Fj.1523-1755.2000.00982.x},
	year = 2000,
	month = {apr},
	publisher = {Elsevier {BV}},
	volume = {57},
	number = {4},
	pages = {1404--1407},
	author = {Gordon H. Williams and Naomi D.L. Fisher and Steve C. Hunt and Xavier Jeunemaitre and Paul N. Hopkins and Norman K. Hollenberg},
	title = {Effects of gender and genotype on the phenotypic expression of nonmodulating essential hypertension},
	journal = {Kidney International}
}
 





\end{document}


\journaltitle{Journal Title Here}
\DOI{DOI HERE}
\copyrightyear{2023}
\pubyear{2019}
\access{Advance Access Publication Date: Day Month Year}
\appnotes{Paper}

\firstpage{1}


\title[Identifying genes associated with phenotype]{Identifying genes associated with phenotypes using machine and deep learning}

\author[1,2]{Muhammad Muneeb}
\author[1,2,$\ast$]{David B. Ascher}
\author[1,2]{YooChan Myung}

\authormark{Muneeb et al.}

\address[1]{\orgdiv{School of Chemistry and Molecular Biology}, \orgname{The University of Queensland}, \orgaddress{\street{Queen Street}, \postcode{4067}, \state{Queensland}, \country{Australia}}}
\address[2]{\orgdiv{Computational Biology and Clinical Informatics}, \orgname{Baker Heart and Diabetes Institute}, \orgaddress{\street{Commercial Road}, \postcode{3004}, \state{Victoria}, \country{Australia}}}
 
\corresp[$\ast$]{Corresponding authors: David B. Ascher, Email: \href{email:d.ascher@uq.edu.au}{d.ascher@uq.edu.au}} 


\received{Date}{0}{Year}
\revised{Date}{0}{Year}
\accepted{Date}{0}{Year}



\abstract{ 
Supplementary Material 1. 
}


\maketitle

\section{Supplementary Material 1}
This section describes the dataset, data processing steps, phenotypic value cleaning, and genotype data transformation.

\subsection{Dataset pre-processing}
There were \texttt{6,401} genotype files and \texttt{668} phenotypes in the openSNP dataset \cite{Greshake2014}, and the phenotype data are available on GitHub (Analysis1.pdf). Only binary phenotypes were considered; phenotypes with missing values were discarded, and each value for each phenotype was assigned to one of three values: Case, Control, or Unknown.

\subsection{Phenotype values transformation}
Participants' phenotypic responses varied, necessitating the cleansing of phenotype data. For example, many terms are used for right-handedness: "right-handed", "right", "Right", and "R". To create cases and controls for each phenotype, we aggregated these occurrences into a single phenotypic value. Other phenotypic value ambiguities are shown on GitHub (AmbiguousPhenotypes.pdf) with examples. 

The \texttt{preTransform file}, which contains the phenotypic information, the number of values for each phenotype, and its unique values, must be created first. \texttt{PreTransform.csv} is available on GitHub and aids in identifying ambiguities in the phenotypic file. The next phase was to create the \texttt{postTransform file}, which converts each phenotype's values from multiple values to one of the following for binary classification.

\begin{itemize}
    \item Cases: Yes (Presence of a particular phenotype value). 
    \item Controls: No (Absence of a particular phenotype value). 
    \item U: Unsure, means we are unsure whether people who selected this phenotype fall in Class1 or Class2.
    \item x: means this phenotype value is removed from further analysis.
\end{itemize}
 
 The final postTransformed is available on this GitHub (postTransformed.csv). 

\subsection{Extract samples for each phenotype}
Convert the values in the preTransform file to the postTransform file, count controls and cases for each phenotype, and finally extract the individuals' genetic files. On GitHub (Analysis2.pdf), you can find the phenotypic name, the number of samples from classes 1 and 2, the total number of samples for each phenotype, and the actual class name. For further analysis, we considered 100 phenotypes.

\subsection{Genetic file format conversion}
The genetic data came in a variety of formats. Thus, we transformed each AncestryDNA file into the 23andMe format before converting it to the PLINK format (bed, bim, fam) \cite{9802470}.

\section{Supplementary Material 2}
This section explains the classification results from the machine learning and deep learning models. 

Tables \ref{machineAUC}, \ref{machinef1score}, and \ref{machineMCC} show the best machine learning model, which yielded the best performance in terms of AUC, F1 Score, and MCC. Tables \ref{deepAUC}, \ref{deepf1score}, and \ref{deepMCC} show the best deep learning model, which yielded the best performance in terms of AUC, F1 Score, and MCC.

\begin{table*}[!ht]
\hskip-1.0cm\begin{tabular}{|l|l|l|l|l|}
\hline
\textbf{Phenotype} & \textbf{AUC} & \textbf{SD} & \textbf{Machine learning algorithm index} & \textbf{Number of SNPs} \\ \hline
\textbf{ADHD} & 56.6 & 4.98 & Xgboost, Booster: gblinear - Loss function: binary:hinge & 5000 \\ \hline
\textbf{Allergicrhinitis} & 66.2 & 26.18 & Xgboost, Booster: gblinear - Loss function: binary:hinge & 200 \\ \hline
\textbf{Asthma} & 62.6 & 7.77 & Xgboost, Booster: gblinear - Loss function: binary:hinge & 1000 \\ \hline
\textbf{Bipolar disorder} & 73.2 & 30.36 & Xgboost, Booster: gbtree - Loss function: binary:logitraw & 500 \\ \hline
\textbf{Cholesterol} & 65.4 & 8.93 & Xgboost, Booster: gbtree - Loss function: binary:hinge & 5000 \\ \hline
\textbf{Craves sugar} & 82.8 & 18.34 & Xgboost, Booster: gblinear - Loss function: binary:logistic & 500 \\ \hline
\textbf{Dental decay} & 66.0 & 8.94 & PassiveAggressiveClassifier & 500 \\ \hline
\textbf{Depression} & 66.0 & 20.64 & Xgboost, Booster: gblinear - Loss function: binary:hinge & 10000 \\ \hline
\textbf{Diagnosed vitamin D deficiency} & 60.8 & 14.52 & Xgboost, Booster: gblinear - Loss function: binary:hinge & 200 \\ \hline
\textbf{Diagnosed with sleep apnea} & 67.0 & 8.37 & Xgboost, Booster: gblinear - Loss function: binary:logistic & 100 \\ \hline
\textbf{Dyslexia} & 63.6 & 3.29 & SGDClassifier & 10000 \\ \hline
\textbf{Earlobe free or attached} & 63.2 & 12.56 & RidgeClassifier & 50 \\ \hline
\textbf{Hair type} & 66.4 & 11.06 & Xgboost, Booster: gblinear - Loss function: binary:logistic & 1000 \\ \hline
\textbf{Hypertension} & 79.8 & 21.82 & DecisionTreeClassifier & 500 \\ \hline
\textbf{Hypertriglyceridemia} & 62.4 & 17.5 & PassiveAggressiveClassifier & 50 \\ \hline
\textbf{Irritable bowel syndrome} & 59.8 & 8.07 & Xgboost, Booster: gblinear - Loss function: binary:logistic & 500 \\ \hline
\textbf{Mental disease} & 60.4 & 9.18 & SGDClassifier & 200 \\ \hline
\textbf{Migraine} & 64.2 & 9.34 & GradientBoostingClassifier & 10000 \\ \hline
\textbf{Motion sickness} & 62.6 & 3.78 & Xgboost, Booster: gblinear - Loss function: binary:hinge & 100 \\ \hline
\textbf{Panic disorder} & 65.6 & 5.59 & Xgboost, Booster: gblinear - Loss function: binary:logitraw & 5000 \\ \hline
\textbf{Photic sneeze reflex photoptarmis} & 60.0 & 6.44 & SGDClassifier & 200 \\ \hline
\textbf{PTSD} & 74.0 & 23.02 & Xgboost, Booster: gblinear - Loss function: binary:hinge & 1000 \\ \hline
\textbf{Scoliosis} & 69.8 & 12.6 & Xgboost, Booster: gblinear - Loss function: binary:logitraw & 100 \\ \hline
\textbf{Sensitivity to Mosquito bites} & 61.0 & 13.51 & Xgboost, Booster: gblinear - Loss function: binary:hinge & 5000 \\ \hline
\textbf{Sleep disorders} & 68.4 & 18.64 & SGDClassifier & 200 \\ \hline
\textbf{Strabismus} & 67.8 & 12.66 & SGDClassifier & 1000 \\ \hline
\textbf{Thyroid issues cancer} & 76.0 & 13.87 & Xgboost, Booster: gblinear - Loss function: binary:hinge & 100 \\ \hline
\textbf{TypeIIDiabetes} & 69.6 & 15.65 & PassiveAggressiveClassifier & 500 \\ \hline
\textbf{Eczema} & 83.2 & 37.57 & Xgboost, Booster: gbtree - Loss function: binary:hinge & 100 \\ \hline
\textbf{Restless leg syndrome} & 66.0 & 17.68 & AdaBoostClassifier & 5000 \\ \hline
\end{tabular}
\caption{\textbf{A table showing the best-performing machine learning algorithm in terms of AUC.} The first column shows the phenotype, the second column shows the average AUC for five-fold, the third column shows the standard deviation, the fourth column shows the machine learning algorithm and hyperparameter that yielded the best performance for a particular phenotype, and the fifth column shows the number of SNPs yielding the best performance.}
\label{machineAUC}

\end{table*}

\begin{table*}[!ht]
\begin{tabular}{|l|l|l|l|l|}
\hline
\textbf{Phenotype} & \textbf{F1 Score} & \textbf{SD} & \textbf{Machine learning algorithm index} & \textbf{Number of SNPs} \\ \hline
\textbf{ADHD} & 45.4 & 8.26 & SGDClassifier & 1000 \\ \hline
\textbf{Allergicrhinitis} & 11.4 & 6.77 & SGDClassifier & 10000 \\ \hline
\textbf{Asthma} & 37.4 & 14.52 & PassiveAggressiveClassifier & 100 \\ \hline
\textbf{Bipolar disorder} & 88.0 & 6.71 & ExtraTreesClassifier & 500 \\ \hline
\textbf{Cholesterol} & 51.8 & 17.06 & DecisionTreeClassifier & 10000 \\ \hline
\textbf{Craves sugar} & 92.0 & 0.0 & BaggingClassifier & 200 \\ \hline
\textbf{Dental decay} & 90.0 & 0.0 & BaggingClassifier & 50 \\ \hline
\textbf{Depression} & 37.4 & 10.99 & SGDClassifier & 10000 \\ \hline
\textbf{Diagnosed vitamin D deficiency} & 38.0 & 14.88 & SGDClassifier & 1000 \\ \hline
\textbf{Diagnosed with sleep apnea} & 55.8 & 5.36 & PassiveAggressiveClassifier & 100 \\ \hline
\textbf{Dyslexia} & 30.0 & 1.87 & SGDClassifier & 10000 \\ \hline
\textbf{Earlobe free or attached} & 43.4 & 13.03 & SGDClassifier & 500 \\ \hline
\textbf{Hair type} & 38.0 & 17.39 & SGDClassifier & 100 \\ \hline
\textbf{Hypertension} & 91.0 & 8.22 & ExtraTreesClassifier & 5000 \\ \hline
\textbf{Hypertriglyceridemia} & 64.6 & 16.61 & PassiveAggressiveClassifier & 50 \\ \hline
\textbf{Irritable bowel syndrome} & 48.2 & 7.26 & SGDClassifier & 200 \\ \hline
\textbf{Mental disease} & 49.2 & 9.86 & SGDClassifier & 200 \\ \hline
\textbf{Migraine} & 71.2 & 2.39 & RidgeClassifier & 10000 \\ \hline
\textbf{Motion sickness} & 61.0 & 5.83 & PassiveAggressiveClassifier & 10000 \\ \hline
\textbf{Panic disorder} & 45.2 & 7.19 & SGDClassifier & 10000 \\ \hline
\textbf{Photic sneeze reflex photoptarmis} & 61.0 & 6.4 & SGDClassifier & 200 \\ \hline
\textbf{PTSD} & 84.4 & 3.13 & BaggingClassifier & 500 \\ \hline
\textbf{Scoliosis} & 57.8 & 8.01 & SGDClassifier & 5000 \\ \hline
\textbf{Sensitivity to Mosquito bites} & 46.4 & 7.8 & SGDClassifier & 500 \\ \hline
\textbf{Sleep disorders} & 36.8 & 17.46 & SGDClassifier & 200 \\ \hline
\textbf{Strabismus} & 38.8 & 11.01 & SGDClassifier & 1000 \\ \hline
\textbf{Thyroid issues cancer} & 63.8 & 19.32 & RidgeClassifier & 100 \\ \hline
\textbf{TypeIIDiabetes} & 36.2 & 15.17 & PassiveAggressiveClassifier & 500 \\ \hline
\textbf{Eczema} & 71.2 & 27.88 & DecisionTreeClassifier & 500 \\ \hline
\textbf{Restless leg syndrome} & 46.0 & 28.81 & AdaBoostClassifier & 5000 \\ \hline
\end{tabular}
\caption{\textbf{A table showing the best-performing machine learning algorithm in terms of F1-Score.} The first column shows the phenotype, the second column shows the average F1-Score for five-folds, the third column shows the standard deviation, the fourth column shows the machine learning algorithm and hyperparameter that yielded the best performance for a particular phenotype, and the fifth column shows the number of SNPs yielding the best performance.}
\label{machinef1score}
\end{table*}

\begin{table*}[!ht]
\begin{tabular}{|l|l|l|l|l|}
\hline
\textbf{Phenotype} & \textbf{MCC} & \textbf{SD} & \textbf{Machine learning algorithm index} & \textbf{Number of SNPs} \\ \hline
\textbf{ADHD} & 12.2 & 14.52 & SGDClassifier & 500 \\ \hline
\textbf{Allergicrhinitis} & 0.0 & 0.0 & AdaBoostClassifier & 200 \\ \hline
\textbf{Asthma} & 15.6 & 24.58 & PassiveAggressiveClassifier & 100 \\ \hline
\textbf{Bipolar disorder} & 36.2 & 52.52 & BaggingClassifier & 10000 \\ \hline
\textbf{Cholesterol} & 25.6 & 30.92 & DecisionTreeClassifier & 10000 \\ \hline
\textbf{Craves sugar} & 31.8 & 37.42 & PassiveAggressiveClassifier & 50 \\ \hline
\textbf{Dental decay} & 24.8 & 13.86 & PassiveAggressiveClassifier & 500 \\ \hline
\textbf{Depression} & 22.6 & 10.85 & DecisionTreeClassifier & 50 \\ \hline
\textbf{Diagnosed vitamin D deficiency} & 13.2 & 29.52 & BaggingClassifier & 100 \\ \hline
\textbf{Diagnosed with sleep apnea} & 30.0 & 30.31 & AdaBoostClassifier & 500 \\ \hline
\textbf{Dyslexia} & 19.0 & 4.69 & SGDClassifier & 10000 \\ \hline
\textbf{Earlobe free or attached} & 29.6 & 29.57 & RidgeClassifier & 50 \\ \hline
\textbf{Hair type} & 25.6 & 14.31 & BernoulliNB & 10000 \\ \hline
\textbf{Hypertension} & 58.0 & 43.35 & DecisionTreeClassifier & 500 \\ \hline
\textbf{Hypertriglyceridemia} & 25.4 & 36.07 & PassiveAggressiveClassifier & 50 \\ \hline
\textbf{Irritable bowel syndrome} & 15.8 & 19.21 & BernoulliNB & 500 \\ \hline
\textbf{Mental disease} & 20.4 & 18.45 & SGDClassifier & 200 \\ \hline
\textbf{Migraine} & 29.0 & 18.75 & GradientBoostingClassifier & 10000 \\ \hline
\textbf{Motion sickness} & 23.8 & 22.88 & SGDClassifier & 200 \\ \hline
\textbf{Panic disorder} & 13.0 & 19.14 & GradientBoostingClassifier & 500 \\ \hline
\textbf{Photic sneeze reflex photoptarmis} & 21.0 & 12.88 & SGDClassifier & 200 \\ \hline
\textbf{PTSD} & 12.8 & 28.62 & BaggingClassifier & 500 \\ \hline
\textbf{Scoliosis} & 32.2 & 54.17 & RandomForestClassifier & 100 \\ \hline
\textbf{Sensitivity to Mosquito bites} & 12.4 & 17.34 & SGDClassifier & 500 \\ \hline
\textbf{Sleep disorders} & 25.2 & 26.98 & SGDClassifier & 200 \\ \hline
\textbf{Strabismus} & 26.2 & 18.59 & SGDClassifier & 1000 \\ \hline
\textbf{Thyroid issues cancer} & 35.4 & 39.73 & RidgeClassifier & 100 \\ \hline
\textbf{TypeIIDiabetes} & 26.8 & 22.7 & PassiveAggressiveClassifier & 500 \\ \hline
\textbf{Eczema} & 52.2 & 51.5 & DecisionTreeClassifier & 500 \\ \hline
\textbf{Restless leg syndrome} & 30.4 & 33.78 & AdaBoostClassifier & 5000 \\ \hline
\end{tabular}
\caption{\textbf{A table showing the best-performing machine learning algorithm in terms of MCC.} The first column shows the phenotype, the second column shows the average MCC for five-folds, the third column shows the standard deviation, the fourth column shows the machine learning algorithm and hyperparameter that yielded the best performance for a particular phenotype, and the fifth column shows the number of SNPs yielding the best performance.}
\label{machineMCC}
\end{table*}

\begin{table*}[!ht]
\hskip-1.0cm\begin{tabular}{|l|l|l|l|l|}
\hline
\textbf{Phenotype} & \textbf{AUC} & \textbf{SD} & \textbf{Algorithm-Dropout-Optimizer-BatchSize-Epochs} & \textbf{Number of SNPs} \\ \hline
\textbf{ADHD} & 55.6 & 9.74 & BILSTM-0.2-Adam-5-200 & 10000 \\ \hline
\textbf{Allergicrhinitis} & 67.4 & 5.86 & ANN-0.2-Adam-5-50 & 500 \\ \hline
\textbf{Asthma} & 63.2 & 12.74 & ANN-0.2-Adam-5-50 & 100 \\ \hline
\textbf{Bipolar disorder} & 73.0 & 14.82 & GRU-0.2-Adam-5-200 & 50 \\ \hline
\textbf{Cholesterol} & 63.8 & 9.5 & Stack GRU - BILSTM-0.2-Adam-1-200 & 10000 \\ \hline
\textbf{Craves sugar} & 78.0 & 26.12 & GRU-0.5-Adam-5-50 & 100 \\ \hline
\textbf{Dental decay} & 72.0 & 4.47 & Stack GRU - BILSTM-0.5-Adam-5-50 & 500 \\ \hline
\textbf{Depression} & 62.2 & 6.1 & ANN-0.2-Adam-1-50 & 50 \\ \hline
\textbf{Diagnosed vitamin D deficiency} & 58.2 & 12.93 & ANN-0.2-Adam-5-50 & 1000 \\ \hline
\textbf{Diagnosed with sleep apnea} & 66.4 & 7.8 & LSTM-0.2-Adam-1-200 & 10000 \\ \hline
\textbf{Dyslexia} & 65.6 & 7.47 & Stack LSTM - GRU-0.5-Adam-5-50 & 5000 \\ \hline
\textbf{Earlobe free or attached} & 64.6 & 8.08 & BILSTM-0.5-Adam-1-50 & 5000 \\ \hline
\textbf{Hair type} & 58.8 & 8.81 & Stack GRU - LSTM-0.2-Adam-1-50 & 1000 \\ \hline
\textbf{Hypertension} & 83.2 & 29.01 & LSTM-0.2-Adam-1-50 & 10000 \\ \hline
\textbf{Hypertriglyceridemia} & 64.8 & 27.12 & ANN-0.2-Adam-1-50 & 50 \\ \hline
\textbf{Irritable bowel syndrome} & 54.8 & 5.63 & ANN-0.5-Adam-5-50 & 200 \\ \hline
\textbf{Mental disease} & 60.0 & 10.17 & Stack LSTM - GRU-0.5-Adam-1-50 & 1000 \\ \hline
\textbf{Migraine} & 56.8 & 7.4 & ANN-0.2-Adam-5-200 & 500 \\ \hline
\textbf{Motion sickness} & 59.6 & 8.59 & ANN-0.5-Adam-5-200 & 5000 \\ \hline
\textbf{Panic disorder} & 57.4 & 10.88 & Stack GRU - LSTM-0.5-Adam-5-200 & 5000 \\ \hline
\textbf{Photic sneeze reflex photoptarmis} & 65.8 & 8.41 & Stack GRU - BILSTM-0.5-Adam-1-200 & 5000 \\ \hline
\textbf{PTSD} & 64.6 & 20.8 & ANN-0.5-Adam-5-50 & 500 \\ \hline
\textbf{Scoliosis} & 69.6 & 22.74 & ANN-0.5-Adam-1-200 & 100 \\ \hline
\textbf{Sensitivity to Mosquito bites} & 68.4 & 12.74 & Stack LSTM - GRU-0.5-Adam-1-50 & 1000 \\ \hline
\textbf{Sleep disorders} & 66.8 & 11.65 & Stack GRU - LSTM-0.5-Adam-1-200 & 100 \\ \hline
\textbf{Strabismus} & 71.0 & 11.07 & ANN-0.2-Adam-1-50 & 500 \\ \hline
\textbf{Thyroid issues cancer} & 71.4 & 14.55 & ANN-0.5-Adam-5-200 & 100 \\ \hline
\textbf{TypeIIDiabetes} & 78.6 & 13.67 & Stack BILSTM - GRU-0.2-Adam-1-50 & 1000 \\ \hline
\textbf{Eczema} & 64.8 & 19.98 & LSTM-0.5-Adam-1-200 & 50 \\ \hline
\textbf{Restless leg syndrome} & 54.8 & 24.0 & ANN-0.2-Adam-5-200 & 5000 \\ \hline
\end{tabular}
\caption{\textbf{A table showing the best-performing deep learning algorithm in terms of AUC.} The first column shows the phenotype, the second column shows the average AUC for five-folds, the third column shows the standard deviation, the fourth column shows the deep learning algorithm and hyperparameter that yielded the best performance for a particular phenotype, and the fifth column shows the number of SNPs yielding the best performance.}
\label{deepAUC}
\end{table*}

\begin{table*}[!ht]
\hskip-1.0cm\begin{tabular}{|l|l|l|l|l|}
\hline
\textbf{Phenotype} & \textbf{F1 Score} & \textbf{SD} & \textbf{Algorithm-Dropout-Optimizer-BatchSize-Epochs} & \textbf{Number of SNPs} \\ \hline
\textbf{ADHD} & 40.6 & 8.38 & ANN-0.5-Adam-1-200 & 200 \\ \hline
\textbf{Allergicrhinitis} & 18.2 & 2.86 & ANN-0.2-Adam-5-50 & 500 \\ \hline
\textbf{Asthma} & 42.2 & 11.71 & ANN-0.2-Adam-5-50 & 100 \\ \hline
\textbf{Bipolar disorder} & 77.4 & 10.41 & Stack GRU - LSTM-0.2-Adam-1-200 & 5000 \\ \hline
\textbf{Cholesterol} & 53.6 & 17.34 & ANN-0.2-Adam-5-200 & 200 \\ \hline
\textbf{Craves sugar} & 89.4 & 3.71 & GRU-0.5-Adam-5-200 & 50 \\ \hline
\textbf{Dental decay} & 78.4 & 12.03 & GRU-0.5-Adam-1-50 & 50 \\ \hline
\textbf{Depression} & 43.0 & 13.29 & Stack LSTM - BILSTM-0.2-Adam-1-50 & 10000 \\ \hline
\textbf{Diagnosed vitamin D deficiency} & 39.6 & 6.07 & Stack LSTM - GRU-0.5-Adam-5-50 & 10000 \\ \hline
\textbf{Diagnosed with sleep apnea} & 61.4 & 6.62 & ANN-0.5-Adam-1-50 & 100 \\ \hline
\textbf{Dyslexia} & 36.8 & 16.16 & GRU-0.5-Adam-5-50 & 10000 \\ \hline
\textbf{Earlobe free or attached} & 46.0 & 8.28 & BILSTM-0.5-Adam-1-50 & 5000 \\ \hline
\textbf{Hair type} & 42.4 & 17.42 & ANN-0.2-Adam-1-50 & 200 \\ \hline
\textbf{Hypertension} & 94.0 & 8.22 & GRU-0.2-Adam-1-50 & 10000 \\ \hline
\textbf{Hypertriglyceridemia} & 67.8 & 23.86 & ANN-0.2-Adam-1-50 & 50 \\ \hline
\textbf{Irritable bowel syndrome} & 51.6 & 5.32 & ANN-0.2-Adam-1-200 & 10000 \\ \hline
\textbf{Mental disease} & 49.0 & 6.96 & ANN-0.2-Adam-1-200 & 10000 \\ \hline
\textbf{Migraine} & 61.0 & 7.65 & ANN-0.5-Adam-5-50 & 50 \\ \hline
\textbf{Motion sickness} & 65.4 & 5.27 & ANN-0.5-Adam-5-200 & 5000 \\ \hline
\textbf{Panic disorder} & 49.8 & 8.04 & ANN-0.5-Adam-1-200 & 10000 \\ \hline
\textbf{Photic sneeze reflex photoptarmis} & 64.4 & 10.88 & Stack GRU - BILSTM-0.5-Adam-1-200 & 5000 \\ \hline
\textbf{PTSD} & 72.0 & 8.83 & BILSTM-0.5-Adam-5-200 & 100 \\ \hline
\textbf{Scoliosis} & 67.2 & 23.86 & ANN-0.5-Adam-1-200 & 100 \\ \hline
\textbf{Sensitivity to Mosquito bites} & 57.6 & 12.2 & Stack LSTM - GRU-0.5-Adam-1-50 & 1000 \\ \hline
\textbf{Sleep disorders} & 34.4 & 9.07 & Stack GRU - LSTM-0.5-Adam-1-200 & 100 \\ \hline
\textbf{Strabismus} & 43.0 & 30.36 & Stack LSTM - GRU-0.2-Adam-1-200 & 50 \\ \hline
\textbf{Thyroid issues cancer} & 65.2 & 17.6 & ANN-0.5-Adam-5-200 & 100 \\ \hline
\textbf{TypeIIDiabetes} & 48.6 & 20.84 & Stack BILSTM - GRU-0.2-Adam-5-50 & 500 \\ \hline
\textbf{Eczema} & 60.2 & 11.48 & ANN-0.5-Adam-5-50 & 10000 \\ \hline
\textbf{Restless leg syndrome} & 42.2 & 17.34 & ANN-0.2-Adam-5-200 & 5000 \\ \hline
\end{tabular}
\caption{\textbf{A table showing the best-performing deep learning algorithm in terms of F1-Score.} The first column shows the phenotype, the second column shows the average F1-Score for five-folds, the third column shows the standard deviation, the fourth column shows the deep learning algorithm and hyperparameter that yielded the best performance for a particular phenotype, and the fifth column shows the number of SNPs yielding the best performance.}
\label{deepf1score}
\end{table*}

\begin{table*}[!ht]
\hskip-1.0cm\begin{tabular}{|l|l|l|l|l|}
\hline
Phenotype & \textbf{MCC} & \textbf{SD} & \textbf{Algorithm-Dropout-Optimizer-BatchSize-Epochs} & \textbf{Number of SNPs} \\ \hline
\textbf{ADHD} & 14.0 & 20.77 & BILSTM-0.2-Adam-5-200 & 10000 \\ \hline
\textbf{Allergicrhinitis} & 18.6 & 4.62 & ANN-0.2-Adam-5-50 & 500 \\ \hline
\textbf{Asthma} & 22.2 & 20.41 & ANN-0.2-Adam-5-50 & 100 \\ \hline
\textbf{Bipolar disorder} & 40.8 & 25.06 & GRU-0.2-Adam-5-200 & 50 \\ \hline
\textbf{Cholesterol} & 27.4 & 18.85 & Stack GRU - BILSTM-0.2-Adam-1-200 & 10000 \\ \hline
\textbf{Craves sugar} & 46.6 & 45.5 & GRU-0.5-Adam-5-50 & 100 \\ \hline
\textbf{Dental decay} & 35.6 & 18.23 & Stack BILSTM - GRU-0.5-Adam-5-50 & 500 \\ \hline
\textbf{Depression} & 23.0 & 28.61 & GRU-0.2-Adam-1-50 & 100 \\ \hline
\textbf{Diagnosed vitamin D deficiency} & 15.0 & 21.9 & ANN-0.2-Adam-5-50 & 1000 \\ \hline
\textbf{Diagnosed with sleep apnea} & 34.0 & 16.03 & LSTM-0.2-Adam-1-200 & 10000 \\ \hline
\textbf{Dyslexia} & 29.4 & 22.39 & GRU-0.5-Adam-5-50 & 10000 \\ \hline
\textbf{Earlobe free or attached} & 25.4 & 14.05 & BILSTM-0.5-Adam-1-50 & 5000 \\ \hline
\textbf{Hair type} & 23.4 & 22.66 & Stack GRU - LSTM-0.2-Adam-1-50 & 1000 \\ \hline
\textbf{Hypertension} & 64.8 & 57.76 & LSTM-0.2-Adam-1-50 & 10000 \\ \hline
\textbf{Hypertriglyceridemia} & 33.2 & 45.66 & Stack GRU - LSTM-0.2-Adam-5-200 & 100 \\ \hline
\textbf{Irritable bowel syndrome} & 10.6 & 11.8 & ANN-0.5-Adam-5-50 & 200 \\ \hline
\textbf{Mental disease} & 19.6 & 18.72 & Stack LSTM - GRU-0.5-Adam-1-50 & 1000 \\ \hline
\textbf{Migraine} & 16.0 & 16.28 & ANN-0.2-Adam-5-200 & 500 \\ \hline
\textbf{Motion sickness} & 20.8 & 17.38 & ANN-0.5-Adam-5-200 & 5000 \\ \hline
\textbf{Panic disorder} & 16.0 & 20.68 & Stack GRU - LSTM-0.5-Adam-5-200 & 5000 \\ \hline
\textbf{Photic sneeze reflex photoptarmis} & 33.0 & 16.12 & Stack GRU - BILSTM-0.5-Adam-1-200 & 5000 \\ \hline
\textbf{PTSD} & 27.8 & 36.89 & ANN-0.5-Adam-5-50 & 500 \\ \hline
\textbf{Scoliosis} & 41.2 & 44.67 & ANN-0.5-Adam-1-200 & 100 \\ \hline
\textbf{Sensitivity to Mosquito bites} & 35.8 & 23.21 & Stack LSTM - GRU-0.5-Adam-1-50 & 1000 \\ \hline
\textbf{Sleep disorders} & 23.4 & 14.42 & Stack GRU - BILSTM-0.5-Adam-1-50 & 100 \\ \hline
\textbf{Strabismus} & 32.6 & 33.78 & BILSTM-0.2-Adam-1-50 & 50 \\ \hline
\textbf{Thyroid issues cancer} & 46.6 & 28.83 & ANN-0.5-Adam-5-200 & 100 \\ \hline
\textbf{TypeIIDiabetes} & 41.8 & 29.47 & Stack BILSTM - GRU-0.2-Adam-5-50 & 500 \\ \hline
\textbf{Eczema} & 32.8 & 45.48 & LSTM-0.5-Adam-1-200 & 50 \\ \hline
\textbf{Restless leg syndrome} & 6.6 & 29.33 & ANN-0.5-Adam-1-200 & 5000 \\ \hline
\end{tabular}
\caption{\textbf{A table showing the best-performing deep learning algorithm in terms of MCC.}  The first column shows the phenotype, the second column shows the average MCC for five-folds, the third column shows the standard deviation, the fourth column shows the deep learning algorithm and hyperparameter that yielded the best performance for a particular phenotype, and the fifth column shows the number of SNPs yielding the best performance.}
\label{deepMCC}
\end{table*}

\clearpage

\section{Supplementary Material 3}
This section lists the common genes identified by the machine and deep learning algorithms and those listed in the GWAS catalog for each phenotype.

\subsection{ADHD}
CUL3, LINC01965, TSNARE1, LRRN2, POC1B, RPL10AP3, PPP1R3B, POLR1H, PPP2R2B, NAV2, MYT1L, CCDC195, DUSP6, RPL6P5, SORCS3, ZMIZ1, TEX41
\subsection{Allergicrhinitis}
PRDM16, AKR1E2, LINC00705, IL1B
\subsection{Asthma}
MED24
\subsection{Bipolardisorder}
MRM2, PRKAG2, RNA5SP403, FOXN2, PPP1R21-DT, RNA5SP404, AHCYP3, RASIP1, "-", NCAN, SVEP1, TCF23
\subsection{Cholesterol}
Y\_RNA, CUX1, CPLX3, RNU6-526P, RNU6-1151P, CEACAM16-AS1, PSRC1, SPOCK3, ABO, SH2B2, CDK6, TRIM31-AS1, SLC22A1, TOMM40, GDNF-AS1, GNAI3, NECTIN2, PPARA, ALDH1A2, RSRP1, RN7SL836P, NKPD1, CETP, GSK3B, MARK4, GIPR, FUT2, ULK3, LIPC-AS1, ABCG8, MPV17L2, LINC02117, CELSR2, LIPC, TRIM31, APOB, LITAF, MACO1
\subsection{Cravessugar}
No gene was identified.
\subsection{Dentaldecay}
CFH, PCDH9
\subsection{Depression}
Y\_RNA, TMPRSS5, SNTN, ZP3P2, ZNF804A, LINC01977, STK19, "-", LINC02033, ZMIZ1, RNA5SP404, RNA5SP403, EHHADH-AS1, KLHDC8B, STK24, ITGA11, SMARCA2, CBX4, CPM, C3orf49, HOMER1, CSMD1, LINC01068, C3orf70, HSPD1P6, DRD2, TLL2, ARNTL, SNX4, ACTG1P22, RN7SL592P, VRK2, USP4
\subsection{DiagnosedVitaminDdeficiency}
ALDH1A2
\subsection{DiagnosedwithSleepApnea}
ACTN3
\subsection{Dyslexia}
FHIT
\subsection{EarlobeFreeorattached}
\subsection{HairType}
TCHH
\subsection{Hypertension}
ALDH1A2, GPR45, PRDM16, LINC02871, CYYR1-AS1, FIGN, MTHFR, PGAM1P5, TGFBRAP1, ATP2B1, FAT1P1, EML6, HIGD1AP3, PRPS1P1, SARM1, "-", MSX2, VTN
\subsection{Hypertriglyceridemia}
No gene was identified.
\subsection{IrritableBowelSyndrome}
No gene was identified.

\subsection{MentalDisease}
LINC02395, TMPRSS5, NTM, POC1B, CHN2, POM121L2, EXOC4, NGF-AS1, LINC01517, KCND2, FOXN2, IL1R2, CPVL, SCN2A, CNIH3, ZSCAN16-AS1, PSMA6P3, ITPRID1, PPP1R21-DT, NCKAP5, IL1R1, NEK6, OR12D3, FARP1, SPTLC1P2, NGF, LINC02151, LINC02326, RIN2, MSH5-SAPCD1, COL6A1, CSMD1, CLU, WDR7, MSH5, ADGRF5, DRD2, SGCG, HMGB1P46, OR5V1, DUSP6, LINC02674, ANK1, LHX2, ACTL8, HIP1
\subsection{Migraine}
No gene was identified.

LINC00645, ASTN2, MIR4307HG, NRP1, MARCHF4, BPIFC
\subsection{Motionsickness}
No gene was identified.

\subsection{PanicDisorder}
No gene was identified.

\subsection{PhoticSneezeReflexPhotoptarmis}
No gene was identified.
\subsection{PosttraumaticStressDisorderorPTSD}
No gene was identified.

\subsection{Scoliosis}
MAGI1
\subsection{SensitivitytoMosquitoBites}
No gene was identified.
\subsection{SleepDisorders}
ASB3, MIR4431
\subsection{Strabismus}
No gene was identified.
\subsection{ThyroidIssuesCancer}
No gene was identified.
\subsection{TypeIIDiabetes}
MTNR1B, CDC123, IGF2BP2, SNRPGP16, KCNJ11, RN7SL198P
\subsection{eczema}
IL13, LINC01991, TH2LCRR, MSMB, BCL6
\subsection{restlesslegsyndrome}
BTBD9

\bibliographystyle{unsrt}
\bibliography{reference}